\input epsf
\newfam\frakfam
\font\teneufm=eufm10   \textfont\frakfam=\teneufm
\font\seveneufm=eufm7   \scriptfont\frakfam=\seveneufm
\font\fiveeufm=eufm5   \scriptscriptfont\frakfam=\fiveeufm

\newfam\msbfam
\font\tenmsb=msbm10   \textfont\msbfam=\tenmsb
\font\sevenmsb=msbm7   \scriptfont\msbfam=\sevenmsb
\font\fivemsb=msbm5   \scriptscriptfont\msbfam=\fivemsb
\def\Bbb{\fam\msbfam \tenmsb}

\magnification\magstep1

\hfuzz=.1in\relax
\def \set#1{\ifmmode \{\,#1\, \}\else $\{\,#1\,\}$\fi}

\def\linadj#1{\normalbaselines
	\multiply\lineskip#1 \divide\lineskip100
	\multiply\baselineskip#1 \divide\baselineskip100
	\multiply\lineskiplimit#1 \divide\lineskiplimit100 }
\def \Im{\mathop{\rm Im}\nolimits}

\def\Re{\mathop{\rm Re}\nolimits}

\newcount \eqnum
\def \chq#1{\global \advance \eqnum1 \xdef#1{(\the \eqnum)}#1}
\def\cheqno{\eqno\chq}

\def \b {\bigskip\noindent}

\def \m {\medskip\noindent}

\def \s {\smallskip\noindent}

\def \pf {\noindent {\bf Proof.}\enskip}

\def \qed {\penalty500\smallskip\penalty500\rightline{QED}\m}

\linadj{140}
\b
\b
\centerline{\bf The Spectrum of the Partially Locked State for the Kuramoto Model }
\b
\centerline{\bf Renato Mirollo* and Steven H. Strogatz$^{\dag}$ }

\vfootnote*{Mathematics Department, Boston College, Chestnut Hill MA 02467 USA}

\vfootnote{$^{\dag}$}{Center for Applied Mathematics and Department of Theoretical and Applied Mechanics, 212 Kimball Hall, Cornell University, Ithaca NY 14853-1503 USA}
\b
\b
\centerline{\it Dedicated to the memory of John David Crawford}
\b
\b
\centerline{\bf Abstract}
\b

We solve a longstanding stability problem for the Kuramoto model of coupled oscillators.  This system has attracted mathematical attention, in part because of its applications in fields ranging from neuroscience to condensed-matter physics, and also because it provides a beautiful connection between nonlinear dynamics and statistical mechanics. The model consists of a large population of phase oscillators with all-to-all sinusoidal coupling.  The oscillators' intrinsic frequencies are randomly distributed across the population according to a prescribed probability density, here taken to be unimodal and symmetric about its mean.  As the coupling between the oscillators is increased, the system spontaneously synchronizes: the oscillators near the center of the frequency distribution lock their phases together and run at the same frequency, while those in the tails remain unlocked and drift at different frequencies.  Although this ``partially locked'' state has been observed in
  simulations for decades, its stability has never been analyzed mathematically.  Part of the difficulty is in formulating a reasonable infinite-$N$ limit of the model.  Here we describe such a continuum limit, and prove that the corresponding partially locked state is, in fact, neutrally stable, contrary to what one might have expected. The possible implications of this result are discussed. \b
\b
\b
\b
\b
\b
\b
Abbreviated title: Spectrum of Partial Locking in the Kuramoto Model
\b
\b
\vfill\eject
\b
\b
\b
\b\noindent{\bf 1. Introduction}
\b

Collective synchronization occurs throughout the living world, from the rhythmic firing of thousands of pacemaker cells in our hearts, to the chorusing of crickets on a warm summer evening [Winfree 1967, 1980; Pikovsky et al.~2001; Strogatz 2003].  What is remarkable is that these and many other biological populations somehow manage to synchronize themselves spontaneously, without any external cue, despite the inevitable diversity in the natural frequencies of their constituent oscillators.

Thirty years ago, Kuramoto introduced an elegant model of such self-synchronizing systems [Kuramoto 1975, 1984; for reviews, see Strogatz 2000 and Acebron et al.~2005].  Although the model was originally inspired by biology [Winfree 1967], it has since found application to many other parts of science and technology.  Examples include the mutual synchronization of electrochemical oscillators [Kiss et al.~2002], metronomes [Pantaleone 2002], Josephson junction arrays [Wiesenfeld, Colet and Strogatz 1996], neutrino flavor oscillations [Pantaleone 1998], collective atomic recoil lasing [von Cube et al.~2004], audiences clapping in unison [Neda  2000], and crowds walking on wobbly footbridges [Strogatz et al.~2005].

Aside from its scientific applications, the model has also been an object of mathematical interest.  Its main virtue has always been its tractability.  In the limit of an infinite number of oscillators, one could ``solve the model exactly,'' in the physicists' sense, as long as one was willing to make some plausible assumptions about the stability and convergence properties of the solutions.  Putting these assumptions on a more rigorous footing has, however, turned out to be problematic.

Indeed, Kuramoto himself realized this from the start and was frank about it.  For instance, in his 1984 book he presents an ingenious formal calculation and then draws attention to its limitations.  Specifically, he shows that as the coupling between the oscillators is increased, the zero solution (corresponding to a completely desynchronized state) bifurcates supercritically to a nonzero solution (corresponding to a partially synchronized state) at a critical value of the coupling strength. He then remarks that the zero solution should be stable below threshold and unstable above it, but writes ``Surprisingly enough, this seemingly obvious fact seems difficult to prove" [Kuramoto 1984, p.~74]. Similarly, he points out that the bifurcating solution is expected to be stable above threshold, though ``Again, this fact appears to be difficult to prove" [Kuramoto 1984, p.~75].

In this paper we settle the second of these issues, namely, the stability of the partially synchronized state.  We find that this state is linearly {\it neutrally} stable, rather than asymptotically stable.  This result may seem puzzling, but there is a precedent for it: the same neutral stability was already established fifteen years ago for the zero solution (now known as the incoherent state) for coupling strengths below the synchronization threshold [Strogatz and Mirollo 1991].

The question studied here may be of interest to readers working on stability analyses in other parts of nonlinear science, wherever continuity equations arise, such as kinetic theory, traffic flow, plasma physics, and fluid mechanics.  The problem formulation involves a nonlinear partial integro-differential equation, one of whose stationary solutions (the partially locked state) contains both a smooth piece and a delta-function piece. To make sense of this, we need to work in an appropriate functional-analytic setting, and carry out the linear stability analysis in a space of suitable ``generalized functions." 

The resulting technical issues are new, at least in this context.  They certainly did not arise in previous studies of the other stationary states of the Kuramoto model.   For example, the stability of the incoherent state can be determined with standard methods, at both the linear [Strogatz and Mirollo 1991, Crawford 1994] and weakly nonlinear [Bonilla et al.~1992, Crawford 1994] levels.  The problem is relatively straightforward because the incoherent state is described by a smooth (in fact, constant) density of oscillators in phase space.  The fully locked state is similarly amenable to conventional techniques, as long as $N$ is finite.  Its stability analysis can be handled with linear algebra [Aeyels and Rogge 2004, Mirollo and Strogatz 2005] or Lyapunov functions [van Hemmen and Wreszinski 1993, Jadbabaie et al.~2004, Chopra and Spong 2006], since the finite-$N$ locked state corresponds to a fixed point for an ordinary differential equation.  Even the partially locked state is susceptible to familiar approaches, if one regularizes the Kuramoto model by adding noise terms to it [Sakaguchi 1988]; then the stability of partial locking at onset follows from the weakly nonlinear analyses mentioned above [Bonilla et al.~1992, Crawford 1994]. 

But none of these simplifications are available for the problem studied here.  Its thornier aspects stem from the combination of a continuum limit, the absence of noise, the need to work far from the onset of instability, and the singular nature of the partially locked state itself.  We imagine that a similar mix of ingredients could crop up in stability problems in other parts of nonlinear science, and hence may be of wider interest. 

The goal of this paper is threefold: set up the continuum limit of the Kuramoto model in a mathematically precise fashion; describe the fixed states for this model; and carry out the linear stability analysis at these states.  The third of these is by far the most interesting to us, since it has potential to shed light on the still poorly understood dynamics of the finite-$N$ system [Strogatz 2000, Balmforth and Sassi 2000, Maistrenko et al.~2005].  Ultimately, we will achieve a complete understanding of the spectrum of the linearized evolution equation for the fixed states of greatest significance, which we call {\it special positive fixed states}.  These are the only candidates for stability; the other stationary states turn out to be manifestly unstable. 

The organization follows accordingly.  After reviewing the Kuramoto model to establish notation (Section~2), we describe its continuum limit (Section~3) and classify its corresponding fixed states (Section~4).  Included here is the derivation of Kuramoto's original self-consistency equation [Kuramoto 1975, 1984], which becomes rigorous in this setting.  In Section~5 we develop the technical machinery needed to describe the tangent space of the model at a fixed state; this is the natural domain for the linear stability analysis.  We analyze the continuous spectrum of the linearized model in Section~6, and derive a characteristic equation whose roots give us the eigenvalues of the linearization in Section~7.    Finally in Section~8 we prove that the fully locked special positive states are linearly stable, but the partially locked special positive states are only neutrally stable, since the spectrum contains the entire imaginary axis!  The implications of this result for the finite-$N$ model are far from clear, although this vaguely suggests that one should not expect to see any kind of exponential convergence to a stable configuration in the finite-$N$ system in the range of coupling for which there is only partial locking.

Before turning to the analytical development, we would like to add a personal note.  When we began thinking about this stability problem around a decade ago, we found ourselves confused by a number of its features.  As we had done on an earlier occasion [Strogatz et al.~1992], we turned to John David Crawford for advice.  J.~D.~was a brilliant mathematical physicist with expertise in bifurcation theory, plasma physics, and nonlinear science in general.  He was also exceptionally generous and a natural teacher.  We last saw him in the spring of 1998 at a conference on pattern formation at the Institute for Mathematics and its Applications in Minneapolis.  A few years earlier he had been diagnosed with Burkitt's lymphoma, and when we saw him at the meeting, he was frail from chemotherapy but delighted to be able to renew old friendships and to join in the scientific discussions.  In particular, he became curious about the stability problem that is the subject of this paper.  The three of us spent a few afternoons working out some preliminary calculations.  Tragically, J.~D.~passed away later that summer, at age 44.  He was very much in our minds as we gathered the fortitude to finish this project, and we're sure it would have been completed much sooner had J.~D.~ still been on our team.  We are honored to dedicate this work to his memory.

\b
\b
\b\noindent{\bf 2. The Kuramoto model}
\b

The Kuramoto model is the system
$$
\dot \theta_i = \omega_i + { K \over N} \sum_{j=1}^N \sin(\theta_j - \theta_i), \quad i = 1, \dots, N \cheqno\eqK
$$
where $N$ is the number of oscillators, $\theta_i (t)$ is the phase of the $i$th oscillator at time $t$, $\omega_i$ its natural frequency, and $K >0$ the coupling strength.  The right hand side of \eqK\
defines a flow on the $N$-fold torus $T^N$, which is the natural state space for the system.
If the frequencies $\omega_i$ have mean $\overline \omega$, we can go into a moving frame at frequency $\overline \omega$ to transform \eqK\ to a system
where the frequencies have mean $0$.  So we can assume $\overline \omega = 0$ without loss of generality; then fixed points of \eqK\ correspond to phase-locked
solutions in the original reference frame.  We also assume that at least one $\omega_i \ne 0$; otherwise,  \eqK\ is a gradient system and is very easy to analyze. 

To characterize the macroscopic state of the system, it is convenient to introduce a complex order parameter defined by
$$
R e^{i \psi} = { 1 \over N} \sum_{j=1}^N e^{i\theta_j}.
$$
If we think of a state $(\theta_1, \dots, \theta_N)$ as an ordered set of $N$ points $ e^{i\theta_j}$ on the unit circle in the complex plane, then $R e^{i \psi}$
is just the centroid of this configuration.  The radius $0 \leq R \leq 1$ measures the phase coherence of the oscillators and $\psi$ indicates their collective phase.

Using the order parameter, we can rewrite the governing equations as
$$
\dot \theta_i = \omega_i + K R \sin(\psi - \theta_i), \quad i= 1, \dots, N. \cheqno\eqtheta
$$
For a given set of natural frequencies $\omega_1, \dots, \omega_N$ there exists a locking threshold $K_l$ such that \eqK\ has fixed points (fully locked states) if and only if $K \ge K_l$; furthermore, for $K > K_l$, \eqK\ has a unique stable fixed point up to rotational symmetry, and hence has a unique stable fixed point whose order parameter has angle $\psi = 0$  [Aeyels and Rogge 2004, Mirollo and Strogatz 2005].  Equation~\eqtheta\ shows that $K_l \ge |\omega_i|$ for all $i$, so $K_l$ will  be large if just one of the natural frequencies $\omega_i$ is large.  So if the natural frequencies $\omega_i$ are chosen randomly with respect  to a probability density function $g$ on $\Bbb R$ which has infinite support, then as $N \to \infty$  the system \eqK\ will have no fixed points for most selections of $\omega_i$.

Kuramoto's intuition was that one could still predict the asymptotic behavior of the system \eqK\ for large $N$ in the absence of fixed points.  He guessed that as $N \to \infty$  the order parameter might still settle down to an almost constant value, despite the incessant motion of the unlocked oscillators.  Seeking such statistically steady solutions, one can assume the order parameter actually {\it is} a constant $R > 0$ and proceed from there.  Then the oscillators divide into two classes, the locked and drifting oscillators, according to whether equation \eqtheta\ has a fixed point or not; the locked oscillators have natural frequencies $\omega_i \in [-KR,KR]$, whereas the drifting oscillators have $|\omega_i| > KR$.  We call these kind of states {\it partially locked}, assuming there are in fact some drifting oscillators (otherwise we say the state is {\it fully locked}).  Kuramoto showed that on average the drifting oscillators make no contribution to the order parameter $R$.  Then, by computing the contribution from the locked oscillators, he produced a self-consistency equation for $R$.  The $N \to \infty$ limit of this equation has a solution $R > 0$ if and only if $K$ is larger than some critical coupling $K_c$, which Kuramoto computed in terms of the density function $g$.   Numerical simulations later confirmed that the size of the order parameter for the system \eqK\ for large $N$ remains close to the value of $R$ predicted by Kuramoto's self-consistency equation [Sakaguchi and Kuramoto 1986].

It's important to understand that Kuramoto's self-consistency equation is only a heuristic (albeit deeply insightful) calculation, so unfortunately no precise conclusions about the finite-$N$ system can be inferred from it.  However, one can introduce an infinite-$N$ analogue of Kuramoto's  system, which has the advantage that the states analogous to the partially locked configurations described above are fixed states in the infinite-$N$ model.  We replace the oscillators $\theta_i$ and natural frequencies $\omega_i$ with  probability measures $\rho_\omega$, which we think of as describing the distribution of the oscillators with  natural frequency $\omega$ on the circle $S^1$.
 Here $\omega$ ranges over the support of a density function $g$, so a state of the infinite-$N$ Kuramoto model is in effect  a family of probability measures parametrized by the natural frequencies $\omega$.  The measures $\rho_\omega$ evolve according to an evolution
equation motivated by the conservation of oscillators; this is
a continuity equation, or equivalently, a Fokker-Planck
equation with no second-order (diffusion) term.  In this setting, Kuramoto's heuristic calculation can be made perfectly rigorous.

\b
\b
\noindent{\bf 3. The Infinite-$N$ Kuramoto Model}
\b

We now describe the infinite-$N$ Kuramoto model.  Let $\Omega = [-1,1]$ or $\Bbb R$, and let $g(\omega)$ be a probability density function  on $\Omega$, which we think of as specifying a distribution of natural frequencies.  We assume that $g(-\omega) = g(\omega)$;  $g$ is non-increasing on $[0, \infty) \cap \Omega$; and $g$ is continuous on $\Omega$ and nonzero on the interior of $\Omega$.  Two familiar examples are the uniform density given by the constant function $1/2$ on $[-1,1]$, and the standard normal density function.   (For convenience, we extend $g$ to be $0$ outside $\Omega$ in the case $\Omega = [-1,1]$.)  As we shall see, these conditions on $g$ are necessary to facilitate many of the calculations undertaken in this paper. 

Let $Pr(S^1)$ be the space of Borel probability measures on the unit circle.  A state for the model will be a family $\rho_\omega \in Pr(S^1)$,  parametrized by $\omega \in \Omega$.  The map $\omega \mapsto \rho_\omega$ must satisfy at least a mild regularity condition, and to make sense of this we need to put a topology on the space $Pr(S^1)$.
There are various ways of doing this; we choose the one best suited to our purposes.

Consider the Banach space $C^k(S^1)$ of $k$-times continuously differentiable real-valued functions on the circle, where $k$ is a non-negative integer (if $k=0$ then $C^0(S^1) = C(S^1)$, the space of continuous functions on $S^1$).  The norm on $C^k(S^1)$ can be defined by
$$
\| \phi \|_k = \max_{\theta \in S^1} \left( |\phi(\theta)| + |\phi'(\theta)| + \cdots+ |\phi^{[k]}(\theta)| \right)
$$
 for $\phi \in C^k(S^1)$.
 We'll be working with the dual spaces $C^k(S^1)^\ast$ throughout this paper, so it will be helpful to describe their elements as concretely as possible.  Any $\nu \in C^k(S^1)^\ast$ can be represented as follows:
 $$
 \langle \phi, \nu \rangle = \int_{S^1} \phi \, d \mu_0 +\int_{S^1} \phi'  d \mu_1 +  \cdots + \int_{S^1} \phi^{[k]} d \mu_k
$$
where $\mu_0, \dots , \mu_k$ are signed Borel measures on $S^1$ and $\phi$ is any $C^k$ function on $S^1$
(the signed measures $\mu_0, \dots, \mu_k$ are not unique).  We can express this more succinctly as
$$
 \langle \phi, \nu \rangle = \langle1,\nu \rangle  \int_{S^1} \phi \, dm + \int_{S^1} \phi^{[k]} d \mu \cheqno\distdef
$$
where $m$ is normalized Lebesgue measure  and $\mu$ a signed Borel measure with $\mu(S^1) = 0$, which is now uniquely determined by $\nu$.  The elements of $C^k(S^1)^\ast$ can be thought of as a certain class of generalized functions or distributions on $S^1$, which we call {\it $k$th-order distributions};  these are just measures and their first $k$ derivatives, in the sense of distributions.  We can interpret $\nu =  \langle1,\nu \rangle m + (-1)^kD^k \mu$ in this sense.

In particular, a Borel probability measure $\mu$ on $S^1$ is naturally an element of the dual space $C^k(S^1)^\ast$, with the pairing given by integration:
 $$
 \langle \phi, \mu \rangle = \int_{S^1} \phi \, d \mu .
 $$
 This gives an embedding of $Pr(S^1)$ in $C^k(S^1)^\ast$, and we use the dual norm on $C^k(S^1)^\ast$ to induce a metric on $Pr(S^1)$.  A distribution $\nu \in C^k(S^1)^\ast$ is a probability measure if and only if $\langle 1, \nu \rangle = 1$, and $\langle \phi , \nu \rangle \ge 0$ for any $\phi \in C^k(S^1)$ such that $\phi \ge 0$.  This shows that $Pr (S^1)$ is closed in $C^k(S^1)^\ast$ for all $k \ge 0$.  The inclusion map $i: C^k(S^1) \to C(S^1)$ is a compact operator when $k \ge 1$, and hence so is its adjoint $i^\ast : C(S^1)^\ast \to C^k(S^1)^\ast$.  Any probability measure has norm $1$ when considered as an element of $C(S^1)^\ast$,  so $Pr (S^1)$ is contained in the image of the unit ball under the map $i^\ast$, and hence $Pr (S^1)$ is a compact subset of $C^k(S^1)^\ast$.

{\it From now on we insist that $k \ge 1$.}  The compactness of $Pr(S^1)$ in $C^k(S^1)^\ast$ has some desirable consequences.  A compact Hausdorff topology cannot be strengthened without sacrificing compactness, or weakened without sacrificing the Hausdorff property.  This implies that the topology on $Pr(S^1)$ is the same for all $k$.  Furthermore, the so-called weak$^\ast$-topologies on $C^k(S^1)^\ast$ all induce the same topology on $Pr(S^1)$.  The closure of the span of $Pr(S^1)$ in $C^k(S^1)^\ast$ is the subspace  of elements $\nu  \in C^k(S^1)^\ast$ that can be represented in the form \distdef\ with $\mu$ absolutely continuous w.r.t.~Lebesgue measure; we call this subspace $C^k(S^1)^\ast_{abs}$ .  The space $C^k(S^1)^\ast_{abs}$ is separable (it's in fact  isomorphic to $L^1(S^1)$), and so is an ideal choice for a Banach space in which to embed $Pr(S^1)$. (The larger Banach spaces $C^k(S^1)^\ast$ are not separable for all $k \ge 0$.)

Now that we have a good topology on $Pr(S^1)$, we can officially define the states of the infinite-$N$ Kuramoto model.  We need a regularity condition on the states which will allow us to integrate various things; the mildest form of this is the requirement of measurability.  
So we define the states  as follows:

\proclaim Definition.  A state for the infinite-$N$ Kuramoto model is a measurable map $\omega \mapsto \rho_\omega$ from $\Omega$  to $Pr(S^1)$.  We denote the space of states by $\cal S$.

\noindent As is the usual practice, we shall identify  two states which agree for almost all $\omega \in \Omega$, so a state is actually an equivalence class of maps under this relation, but we will usually be tacit about this technicality.  Since the weak$^\ast$ and dual norm  topologies on $C^k(S^1)^\ast$ induce the same topology  on $Pr(S^1)$, the measurability condition is equivalent to requiring that for any $C^k$ function $\phi$ on $S^1$, the function
$$
\omega \mapsto \int_{S^1} \phi \, d \rho_\omega
$$
is  measurable on $\Omega$.
The state space $\cal S$ is naturally a closed subset of the Banach space $ L^1 ( \Omega, C^k(S^1)^\ast_{abs})$  of (equivalence classes of) measurable functions from $\Omega$  to  $C^k(S^1)^\ast_{abs}$ that are integrable with respect to the measure $g(\omega)  d\omega$.   
(See Lang [1993, Chapter VI] for background information on integration of functions with values in a Banach space).

Let $K > 0$ be a constant, which we think of as determining the coupling strength for the model.  The rest of the ingredients in the Kuramoto model can be defined as follows.

\proclaim Definition. Given a state $\rho \in \cal S$,  its order parameter is the complex number
$$
Re^{i \psi} = \int_\Omega \left ( \int_0^{2\pi} e^{i \theta}  d\rho_\omega (\theta) \right) g(\omega)  d\omega. \cheqno\eqR
$$
The vector field $v_\omega$ associated to $\rho$ is the function $v_\omega$ on $S^1$ given by
$$
 v_\omega (\theta) =   \omega + KR \sin (\psi - \theta).
\cheqno\eqv
$$

\noindent Note that the map
$$
\omega \mapsto  \int_0^{2\pi} e^{i \theta} d\rho_\omega(\theta)
$$
is a bounded measurable function of $\omega$,  and so the order parameter $Re^{i\psi}$ is well-defined.
The vector field varies with $\omega$, so it's best to think of this as a family of vector fields $v_\omega$ on $S^1$ parametrized by $\omega \in \Bbb R$, just like the measures $\rho_\omega$.  Now we can finally describe the equation that drives the infinite-$N$ Kuramoto model:

\proclaim Definition.  The evolution equation for states $\rho \in \cal S$ is 
$$
{d \over d t}  ( \rho_\omega  )+  D (v_\omega \rho_\omega ) = 0.  \cheqno\eqev
$$

\noindent Here  $D$ is the derivative on $S^1$, interpreted in the sense of distributions:  if $\phi$ is a $C^k$ function on the circle, then the action of $D (v_\omega \rho_\omega ) $ on $\phi$ is given by
$$
\langle \phi, D (v_\omega \rho_\omega) \rangle = - \int_{S^1} \phi'  v_\omega \, d \rho_\omega .
$$
The evolution equation is motivated by the conservation of oscillators with a given frequency $\omega$;  in fact, \eqev\ is exactly the equation that governs the flow of a measure or distribution on $S^1$ corresponding to the flow determined by the vector field $v_\omega$, except for the added twist that $v_\omega$ depends on $\rho$ through the order parameter, and so \eqev\ is nonlinear.  It can be shown that there exists a unique solution to \eqev\ for any  initial state $\rho$, defined for all time $t$.  
This is not a trivial matter, since the operator  $D$ is unbounded on the space $C^k(S^1)^\ast$, and so the usual existence and uniqueness theorems for ODEs on Banach spaces don't apply.  We present  a brief outline of the proof of this claim, and save the details for a future work devoted to the analysis of the nonlinear system defined by \eqev.

Let $z(t)$ be a continuous function on $[0,T]$ taking values in the closed unit disc, and let 
$$
w_\omega (\theta) = \omega + K \Im (z(t) e^{-i\theta}).
$$
Replace $v_\omega$ by $w_\omega$ in \eqev\  to define an {\it uncoupled} version of \eqev.  For a given initial state $\rho_\omega$, we solve this uncoupled equation for each $\omega$ to obtain a $1$-parameter family $ t \mapsto \rho_{\omega,t}$ and substitute back in \eqR\ to obtain a new continuous function $Z(t)$ on $[0,T]$.  This defines a map $\Phi : z \mapsto Z$ on the space of continuous functions on $[0,T]$ taking values in the unit disc.  Fixed points of $\Phi$ correspond to solutions of the original evolution equation \eqev.  For $T$ sufficiently small (independent of the initial condition), $\Phi$ is a contraction map and so has a unique fixed point; this proves existence and uniqueness of solutions on $[0,T]$, which in turn implies existence and uniqueness for all time $t$.

\b
\b
\noindent{\bf 4. Fixed States}
\b

Our first task is to determine the fixed states for our model.  A fixed state $\rho$ is just  a solution to the equation
$$
D(v_\omega \rho_\omega ) = 0
$$
(for almost all $\omega$). The distributions $\xi$ on $S^1$ that satisfy $D \xi = 0$ are constant multiples of Lebesgue measure $m$ on $S^1$, which we normalize so that $m(S^1) = 1$. So the fixed states satisfy an equation of the form
$$
v_\omega \rho_\omega = C_\omega m,  \cheqno\eqsta
$$
where $C_\omega$ is some coefficient function depending on $\omega$.
If $\rho$ has order parameter $0$ then its associated vector field reduces to $v_\omega (\theta) = \omega$, so we see that $\rho_\omega = m$ for all $\omega$.  We call this $\rho$ the {\it incoherent state}, as mentioned in the introduction.  Note that $\rho$ does indeed have order parameter $0$, since
$$
 \int_0^{2 \pi} e^{i \theta} d\theta = 0
$$
and so the inner integral in \eqR\ is $0$ for all $\omega$.  Hence the incoherent state is the unique fixed state with order parameter $0$.

Now let's try to understand the fixed states which have nonzero order parameter (there are a lot of them).  If $\rho$ is such a state, then so is the rotated state $\rho^{\theta_0}$ given by
$$
d \rho_\omega ^{\theta_0} (\theta) = d \rho_\omega (\theta - \theta_0),
$$
where $\theta_0$ is any fixed angle.  The order parameters for $\rho ^{\theta_0}$ and $\rho$ are related by the factor $e^{i\theta_0}$, so we can narrow our search to states for which the order parameter has angle $\psi = 0$; in other words, we assume the order parameter is some $R > 0$. 

\proclaim Definition.  The positive fixed states are those fixed states $\rho \in \cal S$ for which the order parameter $R > 0$.

Plugging in $\psi = 0$ gives 
$$
v_\omega (\theta) = \omega -KR \sin \theta.
$$
Let $\omega \in \Omega$.  If $|\omega| > KR$, $v_\omega (\theta) \ne 0$ for all $\theta$, and equation \eqsta\ gives
$$
d\rho_\omega (\theta) = { C_\omega \over \omega -KR \sin \theta} \, d\theta.
$$
Because $\rho_\omega$ is a probability measure, we must have
$$
C_\omega^{-1} =  \int_0^{2\pi} {d \theta \over \omega - KR \sin \theta}. \cheqno\Cdef
$$
This integral can be evaluated to obtain
$$
C_\omega =\pm {1 \over 2\pi} \sqrt {\omega^2 - (KR)^2},
$$
where the $\pm$ is the same as the sign of $\omega$; in other words, $C_\omega$ is an odd function of $\omega$ (this observation will be important later).  Thus $\rho_\omega$ is completely determined for $|\omega| > KR$:
$$
d\rho_\omega (\theta) = {1 \over 2\pi} {\sqrt {\omega^2 - (KR)^2} \over |\omega - KR \sin \theta|} \, d\theta. \cheqno\driftdef
$$
It's helpful to imagine that for these natural frequencies the oscillators are continuously distributed on the circle according to the measure defined above; we call these oscillators or frequencies {\it drifting} for this reason.  By the way, when $K$ is sufficiently large there exist positive fixed states which have $KR \ge 1$ (we'll construct these in a moment).  So when $\Omega = [-1,1]$, there exist fixed states which have no drifting oscillators; we call these states {\it fully locked}.

If $|\omega| \le KR$ we must have $C_\omega = 0$ in \eqsta.  To see this, observe that away from the zeros of $v_\omega$, the measure $\rho_\omega$ is given by
$$
d \rho_\omega (\theta) = {C_\omega \over \omega - KR \sin \theta} \, d\theta,
$$
just as before.  But now the denominator changes sign, so $\rho_\omega$ cannot be a positive measure unless $C_\omega = 0$.  Hence $ \rho_\omega$ must be supported on the roots of $v_\omega$.  Let $\theta_\omega$ be the root that satisfies
$$
\sin \theta_\omega = {\omega \over KR}, \quad -{ \pi \over 2} <  \theta_\omega < {\pi \over 2};
$$
notice that this choice corresponds to the {\it stable} fixed point of the one-dimensional flow on the circle defined by $\dot \theta = v_\omega (\theta) = \omega - KR \sin \theta$, for $|\omega| < KR$ with $R$ regarded as fixed.  The other root is $\theta^*_\omega= \pi - \theta_\omega$, and of course it corresponds to the unstable fixed point of the flow on the circle.  

Thus $\rho_\omega$ is just a sum of point masses at these two points.  Let $w(\omega)$ be the weight of the probability measure $\rho_\omega$ at the point $\theta^*_\omega$.  Our measurability assumption on the states $\rho_\omega$ guarantees that $w$ is a measurable function on $[-KR,KR] \cap \Omega$, taking values in $[0,1]$.   The weight at $\theta_\omega$ is of course $1 - w (\omega)$.  So for $|\omega| \le KR$ we have
$$
\rho_\omega =   \left ( 1 - w ( \omega) \right)  \delta_ {\theta_\omega} +  w (\omega ) \delta_{\theta^*_\omega}, \cheqno\lockdef
$$
where we use $\delta_\theta$ to denote the unit point mass measure at the point $\theta$.  The case we are most interested in is when $w(\omega) = 0$ a.e.; in other words, $\rho_\omega$ is a unit point mass in the right half-plane  (so in the general case, $w$ measures the deviation from these preferred states).  The intuition here is that these special states have the best chance of being stable for the full, infinite-$N$ system, since their locked oscillators are all located at their stable positions, at least with respect to perturbations that don't change the order parameter.  This observation motivates the following:

\proclaim Definition.  The special positive fixed states are those fixed states $\rho \in \cal S$ for which the order parameter $R > 0$ and the weight function $w(\omega) = 0$ almost everywhere.

If $\omega$ is equal to either $KR$ or $-KR$, then the only probability measure $\rho_\omega$ satisfying \eqsta\ is a unit point mass at $\pi \over 2$ or $-{\pi \over 2}$, so the values of $w$  at $\pm KR$ are irrelevant.  Of course technically we don't even have to consider this case, since the state $\rho$ is completely determined if we describe  $\rho_\omega$ for almost all $\omega$. Since the frequencies $|\omega| \le KR$ have measure $\rho_\omega$  concentrated at one or two points,  we call these oscillators or frequencies {\it locked}.  Every positive fixed state has some locked oscillators.  If all oscillators are locked, we say the state $\rho$ is {\it fully locked}; of course, this can only happen when $\Omega = [-1,1]$.  Otherwise, we call $\rho$ {\it partially locked}.

The state $\rho$ satisfies the equation
$$
R = \int_\Omega \left ( \int_0^{2\pi} e^{i \theta} d\rho_\omega (\theta) \right ) g(\omega)  d\omega,
$$
which in terms of real and imaginary parts is equivalent to
$$
R = \int_\Omega \left ( \int_0^{2\pi} \cos \theta \,d\rho_\omega (\theta) \right ) g(\omega) d\omega \quad {\rm and} \quad 0 = \int_\Omega \left ( \int_0^{2 \pi} \sin\theta\,d\rho_\omega (\theta) \right )g(\omega) d\omega. \cheqno\eqcs\
$$
 We split each of these integrals according to whether $|\omega| \ge KR$ or $|\omega| \le KR$.  In the first case we have
$$
\int_0^{2\pi} \cos \theta \, d\rho_\omega (\theta) = C_\omega  \int_0^{2\pi} {\cos \theta \, d \theta \over \omega - KR \sin \theta} = 0
$$
since the integrand has a periodic antiderivative on $S^1$.  We also have
$$
\openup\jot\eqalign
{
 \int_0^{2\pi} \sin \theta \,d\rho_\omega (\theta) &= C_\omega  \int_0^{2\pi} {\sin \theta \, d \theta \over \omega - KR \sin \theta} \cr
 &= {\omega C_\omega \over  KR} \int_0^{2 \pi} \left ( {1 \over \omega -KR \sin \theta} - {1\over \omega} \right) \, d\theta \cr
 &= {\omega -2\pi C_\omega \over   KR}.}
$$
This last term is an odd function of $\omega$, and so when we integrate it against $g(\omega)$ on the set $\{ |\omega| \ge KR \} \cap \Omega$ we get zero, since $g$ is even.  Hence the drifting oscillators make zero contribution to the order parameter, in concordance with Kuramoto's original calculation [Kuramoto 1975, 1984].

The locked oscillators make no contribution to the sine integral since for these frequencies
$$
\int_0^{2\pi} \sin \theta \, d \rho_\omega (\theta) = \sin \theta_\omega = {\omega \over KR},
$$
which is an odd function of $\omega$.  For the cosine integral we have
$$
\openup\jot\eqalign
{
\int_0^{2\pi} \cos \theta \, d \rho_\omega(\theta) &=  \left (1 - w( \omega) \right)   \cos (\theta_\omega) + w(\omega) \cos (\theta^*_\omega) \cr
&= \left ( 1-2w ( \omega )\right)  \sqrt{1- \left ({\omega \over KR}\right)^2}. \cr
}
$$
So $R$ and $w$ satisfy the self-consistency equation
$$
R = \int _{-KR} ^{KR} \left (1- 2w ( \omega )\right)  \sqrt{1- \left ({\omega \over KR}\right)^2} g(\omega) d \omega,
$$
or equivalently
$$
K^{-1} =  \int_{-1}^1 \left (1-2w(KRs)\right )\sqrt {1-s^2}\, g(KRs) ds. \cheqno\eqscgen
$$
On the other hand if we begin with some $R > 0$ and a weight function $w$ that satisfies \eqscgen, then the formulas above define a positive fixed state $\rho$ with order parameter $R$, and so we have a complete description of the positive fixed states. We summarize these results below.

\proclaim Proposition 1.  To every fixed state $\rho \in \cal S$ with order parameter $R > 0$ we associate a measurable weight function $w :\Omega \cap [-KR, KR]   \to [0,1]$ that satisfies the self-consistency equation \eqscgen.  Conversely, given $R > 0$ and a measurable function $w :  \Omega \cap [-KR, KR]  \to [0,1]$ which  satisfy \eqscgen,  the state $\rho$ defined by equations \driftdef\ and \lockdef\ is a positive fixed state with order parameter $R$.

Since most of our arguments treat the locked and drifting frequencies separately, we introduce notation for these sets; let
$$
\Omega^l = \Omega \cap [-KR, KR] \quad {\rm and} \quad \Omega^d = \Omega - \Omega^l.
$$
Now let's look in more detail at the special positive fixed states (where $w=0$);   then \eqscgen\ reduces to the simpler self-consistency equation
$$
K^{-1} =  \int_{-1}^1 \sqrt {1-s^2}\, g(KRs) ds. \cheqno\eqsc
$$
We can parametrize all solutions $(K, R)$ to \eqsc\  with $K$, $R>0$ as follows.  Let $t > 0$, and  let $\rho(t)$ be the state defined by equations \driftdef\ and \lockdef\ where we substitute $t$ for the constant $KR$.  Also define
$$
f(t) =  \int_{-1}^1 \sqrt {1-s^2}\, g(ts) ds = 2 \int_0^1 \sqrt {1-s^2}\, g(ts) ds,
$$
and let 
$$
K = f(t)^{-1},  \quad R = tf(t)
$$ 
for $t \in (0, \infty)$.  Then $KR = t$ and the state $\rho(t)$  is a special positive fixed state with order parameter $R$ for the model with coupling constant  $K$. (These states are all simply related: if we let $\tilde \rho = \rho(1)$, then we have
$$
\rho(t)_\omega = \tilde \rho_{\omega \over t}.
$$
So the states $\rho(t)$ are all identical up to a scaling of the frequencies.)

$f$ is continuous, positive, and non-increasing on $[0, \infty)$.  We have
$$
 f(0) = 2g(0)  \int_0^1 \sqrt {1-s^2} \, ds = {\pi g(0) \over 2}, \quad {\rm and} \quad
\lim_{t \to \infty} f(t) = 0.
$$
Therefore the image $f\left((0, \infty)\right)$ is either  $\left (0,{\pi g(0) \over 2}\right)$ or $\left(0,{\pi g(0) \over 2}\right]$, depending on whether the value ${\pi g(0) \over 2}$ is taken on at some $t>0$.  Hence there is a critical coupling constant given by
$$
K_c = f(0)^{-1} =  {2 \over \pi g(0)}
$$
such that \eqsc\ has solutions if $K > K_c$, but not if $K < K_c$.  This is essentially Kuramoto's derivation of the critical coupling value for his model.

What happens at $K = K_c$?  It depends on the behavior of the density function $g$ near $0$.  Let $\omega_0$ be the largest value of $\omega$ such that $g$ is constant on $[0, \omega]$.  If   $\omega_0 >0$ (in other words, if $g$ is locally constant at $0$) then the function $f$ is constant on $[0, \omega_0]$, so there is a family of solutions to \eqsc\ parametrized by $t \in (0, \omega_0]$ which all have $K = K_c$ (Figure~1(b)).  However there are no solutions to \eqsc\ with $K = K_c$ and $R > 0$ when $\omega_0 = 0$ (Figure~1(a),(c)).

If we rewrite the function $R = tf(t)$ as
 $$
 R = 2\int_0^t \sqrt{1- \left ({\omega \over t}\right)^2} g(\omega) d \omega,
$$
then we see  that $R$ is a strictly increasing function of $t$,   with image $(0,1)$ for $t > 0$.
So if we plot the parametric curve $(K,R) = (f(t)^{-1}, t f(t))$ 
in the $K$-$R$ plane, then we obtain a curve $C$ in the first quadrant which defines $R$ as an increasing, continuous function of $K$ for $K > K_c$, with perhaps a vertical segment at $K = K_c$; $R \to 1$ as $K \to \infty$ (Figure~1).

Now suppose we have a positive solution $(K_0, R_0)$ to \eqscgen\  for a weight function $w$ which is not almost everywhere equal to $0$.   If we set $t = K_0R_0$, then \eqscgen\ shows that $K_0^{-1} < f(t)$, so the point $(K_0,R_0)$ will lie on the hyperbola $KR = K_0R_0$,  in the region below the curve  $C$ and above the $K$-axis.  Conversely, if $(K_0,R_0)$ is in this region then we can construct a positive fixed state with these parameters as follows.  Let  $(K_0', R_0')$ be the point on $C$ that intersects the hyperbola $KR = K_0R_0$.  Take the special positive fixed state with parameters $(K_0', R_0')$ and continuously deform its weight function from $w=0$ to $w=1/2$; the corresponding fixed states' parameters will trace all points below $C$ on the hyperbola $KR = K_0R_0$. 

To summarize, we see that for each point $(K,R)$ on the curve $C$ there corresponds a unique special positive fixed state with parameters $K$ and $R$; this state always has weight  function $w = 0$.  And for each point $(K,R)$ in the region between $C$ and the $K$-axis there exist (actually infinitely many) positive fixed states with those parameters; these states all have weight functions that are not a.e.~equal to $0$.  If $\Omega = [-1,1]$, then the points $(K,R)$ on or above the hyperbola $KR = 1$ correspond to fully locked states, whereas points below this hyperbola correspond to partially locked states (Figure~1(a),(b)).  So in this case there is a second critical coupling constant 
$$
K_l = f(1)^{-1}
$$
such that the model has fully locked states if and only if $K \ge K_l$. (An equivalent formula for the locking threshold $K_l$ was first obtained by Ermentrout [1985].) 

We wish to stress the distinction between $K_l$ and $K_c$ because there seems to be occasional confusion about it in the literature.  To put it intuitively, suppose that $K$ is gradually increased from zero. The system remains completely desynchronized until $K$ reaches $K_c$, at which point the first oscillators begin to phase-lock. Thus $K_c$ marks the onset of partial locking.  With further increases in $K$, more and more drifting oscillators are recruited into the synchronized pack.  When $K$ finally reaches $K_l$, the locking process is complete.  Now all the oscillators run at the same frequency. 

Hence, partial locking begins at $K_c$; full locking begins at $K_l$.  Notice that $K_l \ge K_c$, with equality if and only if $g$ is constant on $[-1,1]$, corresponding to a uniform distribution of natural frequencies. If the support of $g$ is $\Bbb R$, full locking is never achieved, so it is natural to define $K_l = \infty$ in this case (Figure~1(c)).

\b
\b
\noindent{\bf 5.  Linearization at Fixed States}
\b

Our next task is to study the linearization of the evolution equation \eqev\ at a fixed state $\rho$, which we assume is either a positive fixed state or the incoherent state.  Our ultimate goal is to describe the spectrum of this linearization, which if contained completely in the left half plane would establish the asymptotic stability of the fixed state $\rho$ in the nonlinear model. The domain of the linearized model will be the tangent space $T_\rho \cal S$ at $\rho$ of the state space $\cal S$, which is a subspace of the Banach space $L^1(\Omega, C^k(S^1)^\ast_{abs})$.  We recall the relevant definitions.

\proclaim Definition.  Suppose $A$ be a subset of a (real) Banach space $E$, and $p \in A$.  The tangent cone $TC_p A$ to $A$ at $p$ is the set of $x \in E$ for which there exists a function $\gamma: [0, t_0) \to A$ for some $t_0 > 0$ such that $\gamma (0) = p$ and  $\gamma'(0+) = x$; i.e.
$$
\lim_{t \to 0+} \Bigl \|{\gamma(t) - p \over t} - x  \Bigr \| = 0.
$$
The tangent space to $A$ at $p$ is $T_pA = TC_pA \cap (-TC_pA)$.

\noindent $TC_pA$ is the set of one-sided tangent vectors at $p$ to the set $A$, and $T_pA$ is the set of two-sided tangent vectors.  $TC_pA$ and $T_pA$ are always closed subsets of $E$. If $A$ is convex, then $TC_pA = \overline{C_pA}$, where $C_p A$ is the convex cone spanned by $A - p$.  Therefore $TC_pA$ is closed under addition and multiplication by non-negative scalars, and the tangent space $T_pA = TC_pA \cap (-TC_pA)$ is a closed subspace of $E$.

Before tackling the tangent spaces $T_\rho \cal S$, let's look at the simpler case of just one probability measure $\mu$ on $S^1$, and try to understand the tangent space $T_\mu Pr(S^1)$ in $C^k(S^1)^\ast_{abs}$ for some fixed $k$ (it turns out that now we'll need $k \ge 2$ to get everything  we want).    We can explicitly describe these tangent spaces, at least for the two types of measures that occur for the fixed states.  A (two-sided) tangent vector at $\mu$ to $Pr(S^1)$  is just the derivative  at $t=0$ of some function  $\gamma(t) =  \mu_t$  in $Pr(S^1)$ with $\mu = \mu_0$ and $t$ ranging over some interval $(-t_0,t_0)$ in $\Bbb R$. This derivative, if it exists in $C^k(S^1)^\ast$, is the $k$th-order distribution $\eta$ on $S^1$ defined by the rule
$$
\langle \phi, \eta \rangle =\left.  {d \over dt} \int_{S^1} \phi \, d\mu_t \right| _{t = 0}
$$
for any $C^k$ function $\phi$ on $S^1$.    For example, let $\gamma(t) = \delta_t$ be the unit  point mass at the point $t$.  For  $k \ge 2$, the derivative of this function at $t = 0$ is the distribution which assigns to any $C^k$ function $\phi$ its derivative $\phi'(0)$; this distribution is just the distributional derivative $-D \delta_{0}$ (we need $k \ge 2$ to insure that $\gamma$ is strongly differentiable in the sense described in our definition above).

 $Pr(S^1)$ is a subset of the hyperplane in $C^k(S^1)^\ast_{abs}$ defined by $\langle 1, \eta \rangle = 1$, so any $\eta \in T_\mu Pr(S^1)$ must satisfy the linear condition $\langle 1, \eta \rangle = 0$.  This is the only constraint on the tangent space $T_\mu Pr(S^1) \subset C^k(S^1)^\ast_{abs}$ if $\mu$ is given by $d\mu(\theta) = \alpha(\theta)d\theta$ for some smooth {\it positive} function $\alpha$ on $S^1$ (as is the case for the measures $\rho_\omega$ for the drifting frequencies $\omega$). To see this, let $\eta$ be any smooth distribution on $S^1$ such that $\langle 1, \eta \rangle = 0$.  Then  $t\eta + \mu$ is a probability measure  for $t$ is sufficiently small, so $\eta \in T_\mu Pr(S^1)$.  The closure of the space of smooth distributions $\nu$ with $\langle 1, \eta \rangle = 0$ in $C^k(S^1)^\ast$ is the codimension-one subspace of $C^k(S^1)^\ast_{abs}$  defined by $\langle 1,\eta\rangle = 0$, which therefore  must be the tangent space $T_\mu Pr(S^1)$.

On the other hand, suppose that $\gamma(t) = \mu_t  \in Pr(S^1)$ for $t \in (-t_0,t_0)$ has derivative $\eta$ at $t=0$, and $\mu_0$ is supported on some compact  $K \subset S^1$.  If $\phi \ge 0$ is a $C^k$ function which vanishes on $K^c$, then
$$
\int_{S^1} \phi \, d\mu_t \ge 0 \quad {\rm and } \quad\int_{S^1}  \phi \, d\mu_0 = 0.
$$
Taking  the derivative at $t = 0$ shows that $\langle \phi , \eta \rangle = 0$.  This implies that the distribution $\eta$ is $0$ when restricted to  $K^c$; in other words, $\eta$ is also supported on $K$.  Consequently any tangent vector $\eta \in T_\mu Pr(S^1)$ satisfies the condition ${\rm supp } (\eta) \subset {\rm supp } (\mu)$.

The extreme points of the convex set $Pr(S^1)$ are the unit point mass measures $\delta_p$, $p \in S^1$.  As one might expect, the tangent spaces at these points are fairly small: $T_{\delta_p}Pr(S^1)$ is just the one-dimensional space spanned by $D\delta_p$ provided that $k \ge 2$ ($T_{\delta_p}Pr(S^1) = \{0\}$ when $k = 1$, since $D\delta_p \not \in C^1(S^1)^\ast_{abs}$).    To see this, take $p = 0$ and let $\phi$ be any smooth function on $S^1$ with $\phi(0) = \phi'(0) = 0$.  Suppose $\gamma(t) = \mu_t  \in Pr(S^1)$ for $t \in (-t_0,t_0)$ has $\gamma(0) = \delta_0$ and $\gamma'(0) = \eta$.  Construct a non-negative smooth function $\tilde\phi$ such that $\tilde\phi(0) = 0$  and $| \phi| \le \tilde \phi$ (take $\tilde \phi(\theta)$ to be a large multiple of $\sin^2(\theta /2)$, for example).  Then
$$
-\langle\tilde \phi, \mu_t  \rangle \le \langle \phi, \mu_t   \rangle \le \langle \tilde\phi, \mu_t  \rangle
$$
and $\langle \tilde \phi, \mu_0 \rangle = \tilde\phi(0) = 0$.  
$\langle \tilde\phi, \eta \rangle = 0$ by the same argument as above, so $\langle \phi ,\eta \rangle = 0$ as well.
This, together with the facts that $\langle 1, \eta \rangle = 0$ and ${\rm supp}(\eta) \subset \{0\}$, imply $\eta = cD\delta_0$ for some $c$.
Similarly, if $\mu$ is a linear combination of two distinct unit point masses $\delta_p$ and $\delta_q$, then the elements of $T_\mu Pr(S^1)$ are of the form
$$
\eta = c_0 (\delta_p - \delta_q) + c_1 D \delta_p + c_2 D \delta_q
$$
where the $c_i$ are constants.

{\it From now on, let's insist that $k \ge 2$.}  This insures us a decent supply of tangent vectors at the locked states.  Now that we understand the tangent spaces for the types of measures that occur for the fixed states $\rho \in \cal S$, it becomes a relatively straighforward matter to describe the tangent spaces $T_\rho \cal S$ we will be working with.  The state space $\cal S$ embeds as a convex subset of the space  $ L^1 ( \Omega, C^k(S^1)^\ast_{abs})$, and we can generalize the arguments above to prove the folllowing.

\proclaim Proposition 2.  The tangent space $T_\rho{\cal S} \subset  L^1 ( \Omega, C^k(S^1)^\ast_{abs})$ at a fixed state $\rho \in \cal S$ consists of all $\eta \in  L^1 ( \Omega, C^k(S^1)^\ast_{abs})$ such that $\eta_\omega \in T_{\rho_\omega} Pr(S^1)$ for almost all $\omega \in \Omega$.

\pf  In one direction, suppose $\eta \in T_\rho \cal S$.  Then there exists a map  $\gamma: t \mapsto \cal S$ defined on some interval $(-t_0, t_0)$ with $\gamma (0) = \rho$
such that
$$
\lim_{t \to 0} \int_\Omega \left  \| {\gamma(t)_\omega - \rho_\omega \over t } - \eta_\omega\right  \| g(\omega) d \omega  = 0,
$$
where the norm is taken in the space $C^k(S^1)^\ast$.  Let $t_n \in (0, t_0)$ be any sequence converging to $0$.  The (real-valued) functions
$$
\omega \mapsto \left  \| {\gamma(t_n)_\omega - \rho_\omega \over t_n } - \eta_\omega\right  \| 
$$
converge to $0$ in the space $L^1(\Omega, {\Bbb R})$ with respect to the measure $g(\omega) d \omega$; hence by passing to a subsequence if necessary, we can assume that
$$
\lim_{n \to \infty}  \left  \| {\gamma(t_n)_\omega - \rho_\omega  \over t_n } - \eta_\omega\right  \| = 0
$$
for almost all $\omega \in \Omega$.  Hence $\eta_\omega \in \overline {C_{\rho_\omega} Pr(S^1)} = TC_{\rho_\omega}Pr(S^1)$ for almost all $\omega$.  If we choose $t_n < 0$ we get $\eta_\omega \in -TC_{\rho_\omega} Pr(S^1)$, so
$\eta_\omega \in T_{\rho_\omega} Pr(S^1)$ for almost all $\omega$.

Now consider the set  $V = \{ \eta \in L^1 ( \Omega, C^k(S^1)^\ast_{abs}) \ \big| \ \eta_\omega \in T_{\rho_\omega} Pr(S^1) \  {\rm a.e.} \}$; we wish to show that $V \subset T_\rho{\cal S} $.  Both $V$ and $T_\rho{\cal S} $ are subspaces of $ L^1 ( \Omega, C^k(S^1)^\ast_{abs})$ and  $T_\rho{\cal S} $ is  closed, so it suffices to prove that $T_\rho{\cal S} $ contains a set  of elements $\eta \in V$ whose span is dense in $V$.  Any $\eta \in V$ can be expressed as a sum $\eta = \eta^l+\eta^d$, where $\eta^l, \eta^d \in V$ are supported on the locked and drifting frequencies respectively, so we can consider these cases separately.  The most general $\eta \in V$ supported on the locked frequencies can be expressed as 
$$
\eta_\omega = c_0 (\omega)(\delta_{\theta_\omega} - \delta_{\theta^*_\omega}) + c_1 (\omega)D\delta_{\theta_\omega} + c_2(\omega)  D\delta_{\theta^*_\omega}
$$
for $\omega \in \Omega^l$, where the coefficients $c_i$ are $L^1$-functions of $\omega$ w.r.t.~the measure $g(\omega) d\omega$, and satisfy the constraints $c_0(\omega) = 0$ if $w(\omega) = 0$ or $1$,  $c_2 (\omega)= 0$ if $w(\omega) = 0$ and $c_1(\omega) = 0$ if $w(\omega) =1$.  We consider these three terms separately.

The element $\eta \in V$ defined by $\eta_\omega = c_0 (\omega)(\delta_{\theta_\omega} - \delta_{\theta^*_\omega})$ can be uniformly approximated by linear combinations of simpler elements
$\eta \in V$ defined by $\eta_\omega = \chi_A (\omega) (\delta_{\theta_\omega} - \delta_{\theta^*_\omega})$, where $A \subset \Omega^l$ is measurable ($\chi_A$ denotes the characteristic function of the set $A$).   Because of the constraints on the function $c_0$, we can also assume that for some $\epsilon > 0$, $\epsilon \le w(\omega) \le 1- \epsilon$ for all  $\omega \in A$.   We must show that this $\eta \in T_\rho \cal S$.  Define $\gamma(t)$ by
$$
\gamma (t)_\omega =  \rho_\omega+t  \chi_A (\omega)(  \delta_ {\theta_\omega} - \delta_{\theta^*_\omega});
$$
clearly $\gamma'(0) = \eta$.  Now $\gamma (t)_\omega$ is a probability measure as long as $|t| \le \epsilon$, so $ \eta = \gamma'(0) \in T_\rho \cal S$.

The proof for the other coefficients is similar.  For the $c_1$-term,  it suffices to prove that the element $\eta \in V$ defined by $\eta_\omega = \chi_A( \omega) D \delta_ {\theta_\omega} $ is in $T_\rho\cal S$, where again $A \subset \Omega^l$ is measurable, and this time we assume that for some $\epsilon > 0$, $w(\omega) \le 1-\epsilon$ for all $\omega \in A$.
Define $\gamma(t)$ by
$$
\gamma(t)_\omega = \rho_\omega + \epsilon \chi_A(\omega)( \delta_{\theta_\omega + t} - \delta_{\theta_\omega});
 $$
then $\gamma'(0) = -\epsilon \eta$.   $\gamma (t)_\omega$ is a probability measure for all $t$, and hence $\eta = -\epsilon^{-1} \gamma'(0) \in T_\rho \cal S$. The argument for the $c_2$-term is exactly the same.  Hence  any $\eta \in V$ supported on the locked frequencies is in $T_\rho \cal S$.

Finally, suppose $\eta \in V$ is supported on the drifting frequencies.  $\eta$ is an element of the Banach space $L^1(\Omega^d , C^k(S^1)^\ast_{abs})$, taking values in the codimension-one  subspace $W \subset C^k(S^1)^\ast_{abs}$ consisting of distributions orthogonal to the constant function $1$ on $S^1$.  Therefore $\eta$ can be uniformly approximated by linear combinations of elements in $V$ of the form
$$
\eta_\omega = \chi_A (\omega) \xi,
$$
where $A \subset \Omega^d$ is  measurable and $\xi$ is a smooth distribution in $W$ ($\xi$ has no dependence on $\omega$).  We can also assume that $A$ has positive distance from the boundary frequencies $\pm KR$.  It suffices to prove that this $\eta \in T_\rho\cal S$.  Define $\gamma(t)$ by
$$
\gamma(t)_\omega= \rho_\omega + t \chi_A(\omega) \xi;
$$
clearly $\gamma'(0) = \eta$.  Since $\xi$ is a smooth distribution, $\xi$ is a signed measure on $S^1$ given by $d\xi(\theta) = \alpha (\theta) d\theta$ where $\alpha$ is a smooth function on $S^1$ with integral $0$.  Now recall that for the drifting frequencies, $\rho_\omega$ is the measure given by \driftdef; the coefficient function has minimum value
$$
{1 \over 2\pi} {\sqrt{\omega^2-(KR)^2} \over  |\omega| + KR} ,
$$
which is uniformly bounded away from $0$ for $\omega \in A$ since we assumed $A$ has positive distance from $\pm KR$.  The function $\alpha$ is bounded on $S^1$, so if $|t|$ is sufficiently small, $\gamma(t)_\omega$ is a probability measure for all $\omega$, and hence $\eta = \gamma'(0) \in T_\rho\cal S$.

\qed

Strictly speaking $T_\rho \cal S$ depends on $k$, even though we supress this dependence in the notation.  As a closed subspace of $ L^1 ( \Omega, C^k(S^1)^\ast_{abs})$, $T_\rho\cal S$ is a Banach space in its own right, and is the natural domain for the linearization of the evolution equation \eqev,  which we turn to next.
If we replace $\rho$ by $\rho + \epsilon \eta$ in \eqv, we see that the first-order perturbation of the vector field $v_\omega$ corresponding to a tangent vector  $\eta \in T_\rho \cal S$ is equal to
$$
K\int_{-\infty} ^\infty  \langle \sin (\tau - \theta),  \eta_\omega (\tau) \rangle g(\omega) d\omega = K \left( S\eta \cos \theta -C\eta \sin  \theta \right)
$$
(the pairing inside the integral is with respect to the dummy variable $\tau$).  The coefficients $C\eta$ and $S\eta$ are  given by
$$
\openup\jot\eqalign
{
C\eta &= \int_\Omega  \langle \cos \tau ,  \eta_\omega (\tau) \rangle g(\omega)  d\omega \ \  {\rm and} \cr
S\eta &= \int_\Omega  \langle \sin \tau ,  \eta_\omega (\tau) \rangle g(\omega) d\omega. 
}
$$
$C\eta$ and $S\eta$ are respectively the perturbations of the real and imaginary parts of the order parameter $Re^{i\psi}$ ($C$ for cosine, $S$ for sine).
To linearize  \eqev,  replace $\rho_\omega$ with $\rho_\omega + \epsilon \eta_\omega$ and gather all the linear terms in $\epsilon$ to obtain the equation
$$
{d \over d t} ( \eta_\omega )+  D \bigl(v_\omega  \eta_\omega +    K \left( S\eta \cos \theta -C\eta \sin  \theta \right)
\rho_\omega\bigr) = 0.
$$
This leads to the following definition.

\proclaim Definition.  The linearized evolution equation at a state $\rho \in \cal S$ is  ${d \over dt} (\eta)= L\eta$, where $L$ is  the linear operator defined by
$$
(L \eta) _\omega  = - D\bigl(v_\omega \eta_\omega +  K \left( S\eta \cos \theta -C\eta \sin  \theta \right)
 \rho_\omega \bigr)  \cheqno\eqlin
$$
on the space $T_\rho \cal S$.

The right hand side of \eqlin\ is an integrable family of distributions in $C^{k+1}(S^1)^\ast_{abs}$, but not necessarily in $C^k(S^1)^\ast_{abs}$, since $D$ may map $k$th-order distributions to $(k+1)$st-order distributions.    Hence the operator $L$ is in general unbounded (the exception, as we shall see, is when $\rho$ is fully locked).  However $L$ is a closed, densely-defined operator on the Banach space $T_\rho \cal S$, which is the next best thing to being a bounded operator.

\b
\b
\noindent {\bf  6. The Spectrum of $ \bf L$ for Special Positive States}
\b

The main goal of this paper is to describe completely  the spectrum $\sigma(L)$ for the special positive states.  (For all other positive states, the spectrum  $\sigma(L)$ contains positive numbers, and so these states are linearly unstable, and hence of less interest to us.  We'll comment more on this at the end of this section.)  Since the spectrum may contain complex numbers, we need to study the operator $L$ on the complexified tangent space $T_\rho {\cal S} \otimes \Bbb C$; so from here on we will allow the distributions $\eta_\omega$ to be complex-valued.   We'll denote the complex Banach space $T_\rho {\cal S} \otimes \Bbb C$ simply as $E$.  Following tradition we partition $\sigma (L)$ into three parts: its point, continuous and residual spectrum.  The point spectrum $\sigma _p(L)$ is just the set of eigenvalues of $L$.  The continuous spectrum $\sigma_c(L)$ is the set of $\lambda \in \Bbb C$ such that $\ker (\lambda I - L) = \{0\}$ and the image $\Im (\lambda I - L)$ is dense in $E$, but the inverse $(\lambda I - L)^{-1}$, defined on the dense subspace $\Im (\lambda I - L)$, is unbounded.  Since $L$ is a closed operator, the densely-defined operator  $(\lambda I - L)^{-1}$ is bounded if and only if it is defined on all of $E$, or equivalently, if and only if the image of $\lambda I - L$ is $E$; this is a consequence of the closed graph theorem (see Kato [1995, p.~166]).  So we can also describe $\sigma_c(L)$ as the set of $\lambda$ such that $\lambda I - L$ is one-to-one, has dense image, but is not surjective.  The remainder of the spectrum is the residual spectrum $\sigma_r(L)$, which  is therefore the set of $\lambda \in \Bbb C$ such that $\ker (\lambda I - L) = \{0\}$ and $\Im ( \lambda I - L)$ is contained in a proper closed subspace of $E$.  In this section we begin the analysis of the spectrum of $L$.

{\it Henceforth, unless explicitly noted, we assume that $\rho$ is a special positive state.}  To understand the spectrum of $L$, it helps to express $L = M+ B$, with the operators $M$ and $B$ defined by
$$\openup\jot\eqalign
{
(M \eta) _\omega  &=  - D(v_\omega \eta_\omega), \cr
(B\eta)_\omega &= -K D \bigl( \left(S\eta \cos \theta -C\eta \sin  \theta \right)
 \rho_\omega \bigr) .
}
$$
Notice that $M$ is completely uncoupled, in the sense that $(M\eta)_\omega$ depends only on $\eta_\omega$; in other words, if we define $M_\omega \eta_\omega = - D(v_\omega \eta_\omega)$, then $(M\eta)_\omega = M_\omega \eta_\omega $.  The operator $B$ is  bounded on $E$ provided that $k \ge 2$, which as mentioned earlier is needed to insure that the map $\omega \mapsto D\rho_\omega$  from $\Omega$ to $C^k(S^1)^\ast$ is measurable.  $B$ has a  codimension-two kernel determined by the equations $C\eta = S\eta = 0$, so the rank of $B$ is only 2; in other words, the operators $L$ and $M$ are in a sense very close.  The coupling of the oscillators is entirely expressed through the operator $B$, so one can think of the operator $M$ as describing the linearized Kuramoto model with the coupling artificially suppressed.

We can split the tangent space $E$ as a direct sum $E = E^l \oplus E^d$, where $E^l$ and $E^d$ are the subspaces of tangent vectors supported on the locked and drifting frequencies respectively.  Notice that $M$ preserves both these subspaces.  As we saw in the proof of Proposition 2, $E^l$ is isomorphic to the space of complex-valued functions $L^1(\Omega^l, {\Bbb C})$ (w.r.t.~the measure $g(\omega) d\omega$ on $\Omega^l$).  $E^d$ is the space $L^1(\Omega^d, W)$, where $W \subset C^k(S^1, {\Bbb C})^*_{abs}$ is the subspace of complex-valued distributions orthogonal to the constant function $1$ on $S^1$.
We can also split the tangent space another way, into even and odd subspaces, as follows.  The space $E$ has an involution defined by the rule
$$
(\overline \eta)_\omega (\theta) = \eta_{-\omega} (-\theta).
$$
We call $\eta$ {\it even} if $\overline\eta = \eta$ and {\it odd} if $\overline\eta = -\eta$.  This notion of even and odd behaves much like the usual one for functions of one variable:  any $\eta \in E$ can be expressed as a sum of an even and odd element in a unique way;  if $\eta$ is even then the element $D\eta$ is odd, and vice versa; and if $\phi(\theta)$ is a smooth function on $S^1$, then the usual rules for $\phi \eta$ apply:  if $\phi$ is even (in the traditional sense) and $\eta$ is even, then $\phi\eta$ is even, etc.  Notice that if $\eta \in E$ is even then $S\eta = 0$, since the integrand is an odd function of $\omega$:
$$
  \langle \sin \theta ,  \eta_{-\omega} (\theta) \rangle = \langle \sin \theta ,  \eta_{\omega} (-\theta) \rangle  = \langle \sin (-\theta) ,  \eta_{\omega} (\theta) \rangle = -\langle \sin \theta ,  \eta_{\omega} (\theta) \rangle.
 $$
Similarly, $C\eta = 0$ if $\eta$ is odd.

We denote the even and odd subspaces of $E$ by $E_c$ and $E_s$ respectively, to remind us that the even (odd) tangent vectors correspond to cosine (sine) perturbations of the order parameter.  The even-odd decomposition also respects the locked-drifting decomposition, so we can express $E^l=E^l_c \oplus E^l_s$ and $E^d = E^d_c \oplus E^d_s$ as the direct sum of even and odd subspaces.  Since any even or odd $\eta \in E$ is completely determined by $\eta_\omega$ for $\omega \ge 0$, we can identify
$$
E^l_c \cong E^l_s \cong L^1(\Omega^l_+, {\Bbb C}) \quad {\rm and} \quad E^d_c \cong E^d_s  \cong L^1(\Omega^d_+, W)
$$
where $\Omega^l_+ = \Omega^l \cap [0, \infty)$, $\Omega^d_+ = \Omega^d \cap [0, \infty)$.
The operators $M$ and $L$ both preserve the even and odd subspaces. (Proof:  if $\eta$ is even, then $v_\omega \eta_\omega$ is odd, so $D(v_\omega \eta_\omega)$ is even.  And $(\sin \theta) \rho_\omega(\theta)$ is odd, so $D\left( (\sin \theta) \rho_\omega(\theta)\right)$ is even; hence $M\eta$ and $L\eta$ are even. The proof for $\eta$ odd is similar.)  Hence we can analyze the spectrum of $L$ restricted to the subspaces $E_c$ and $E_s$ separately, and combine the results to obtain the spectrum of $L$ on $E$.

$M$ has only continuous spectrum, and it's fairly easy to describe:

\proclaim Proposition 3.   $\sigma(M) = [-KR, -\sqrt{(KR)^2-1}]$ if $\rho$ is fully locked, and $\sigma(M) = [-KR, 0 ] \cup {\Bbb R}i$  if $\rho$ is partially locked (here ${\Bbb R}i$ denotes the imaginary axis).   All $\lambda \in \sigma(M)$ are in the continuous spectrum.

\noindent The proof of this proposition is somewhat technical, so we postpone it to the end of this section so as not to interrupt the main thread we are developing.   When two closed operators on a Banach space differ by a bounded operator of finite rank, as is the case for $L$ and $M$, their spectra are closely related.  In fact, Proposition 3 is the key ingredient in establishing the main result of this paper.  We present  the proof of this theorem below, with some of the steps deferred to Sections~7 and 8.

\proclaim Theorem.   The spectrum $\sigma(L)$ consists of $\sigma(M)$, an eigenvalue at $0$, and perhaps one other eigenvalue $\lambda \in  [-\sqrt{(KR)^2-1}, 0)$ if $\rho$ is fully locked.  Except for $0$ and $\lambda$, the spectrum of $L$  is all continuous.

\pf  Suppose $\lambda \in \sigma(M)$. The spectrum of $M$ is all continuous, so the image of the operator $\lambda I - M$, which coincides with the domain of the unbounded operator $(\lambda I - M)^{-1}$, must be dense in $E$ and have infinite codimension (otherwise we could extend $(\lambda I - M)^{-1}$ to a closed operator defined on all of $E$ by adding a bounded, finite rank operator, but then the closed graph theorem would imply that $(\lambda I - M)^{-1}$ is bounded on $E$).  Therefore $\lambda I - L$ is also not onto, so $\lambda \in \sigma(L)$; hence $\sigma (M) \subset \sigma(L)$.  Reversing the roles of $L$ and $M$ in this argument shows that $\sigma_c(L) \subset \sigma(M)$.  So to complete the proof we need to prove three things:  
$0$ is always an eigenvalue for $L$ (this is not surprising considering the rotational symmetry of the Kuramoto model); $L$ has at most one other eigenvalue $ \lambda \ne 0$, which satisfies $\lambda \in  [-\sqrt{(KR)^2-1}, 0)$ and occurs only in the fully locked case; and the residual spectrum $\sigma_r(L) = \emptyset$ in all cases.

Let's begin with the eigenvalues.  As discussed above, it suffices to study $L$ separately on the even and odd subspaces $E_c$ and $E_s$; on $E_c$, $L$ is given by
$$
(L \eta) _\omega  = M_\omega \eta_\omega +  K (C\eta) D  \bigl(  (\sin  \theta)
 \rho_\omega \bigr).
$$
$M$  is defined pointwise as a function of $\omega \in \Omega$, and as we shall see in the course of the proof of Proposition 3, for any given $\lambda \in \Bbb C$ the operator $(\lambda I - M_\omega)$ is invertible for all but at most countably many values of $\omega$. So suppose $\eta \in E_c$ is an eigenvector for $\lambda$.  We must have $C\eta \ne 0$, since otherwise $\lambda$ would be an eigenvalue for $M$.  So we can assume that $C\eta = K^{-1}$ if we like.   Then $\eta$ is determined uniquely by the formula
$$
\eta_\omega = (\lambda I - M_\omega)^{-1} D  \bigl(  (\sin  \theta)
 \rho_\omega \bigr).
$$
Conversely if $\lambda\not \in \sigma(M)$ then this equation defines an element $\eta \in E_c$.   But  if $\lambda\in \sigma(M)$ then $\eta$ may not be in $E_c$, since $\eta$ may fail to be integrable  w.r.t.~the density function $g(\omega)$.
$\lambda$ is an eigenvalue for $L$ on $E_c$  if and only if $\eta \in E_c$ and $\lambda$ satisfies the self-consistency relation $C \eta = K^{-1}$.  Define the function $h_c$ by the formula
 $$
 h_c(\lambda) = \int _\Omega \langle \cos \theta, (\lambda I - M_\omega)^{-1} D  \bigl(  (\sin  \theta)
 \rho_\omega \bigr) \rangle g(\omega) d\omega;
 $$
 the domain of $h_c$ is defined to be those $\lambda \in {\Bbb C}$ for which the integrand above is integrable.  Then a necessary condition for $\lambda$ to be an eigenvalue for $L$ on $E_c$ is that $\lambda$ satisfies the self-consistency equation $h_c(\lambda) = K^{-1}$; this condition is necessary and sufficient if $\lambda \not \in \sigma(M)$.

The situation is similar for $E_s$.  Here $L$ is given by
$$
(L \eta) _\omega  = M_\omega \eta_\omega -  K (S\eta) D  \bigl(  (\cos  \theta)
 \rho_\omega \bigr).
$$
We define
 $$
 h_s(\lambda) =- \int _\Omega \langle \sin \theta, (\lambda I - M_\omega)^{-1} D  \bigl(  (\cos  \theta)
 \rho_\omega \bigr) \rangle g(\omega) d\omega;
 $$
 if $\lambda$ is an eigenvalue for $L$ on $E_s$, then $\lambda$ must satisfy the self-consistency equation $h_s(\lambda) = K^{-1}$.  We call $h_c$ and $h_s$ the {\it characteristic functions} for $L$.    We will derive explicit formulas for these functions in the next section, and then in Section~8 we shall prove

 \proclaim Proposition 4.  The equation $h_c(\lambda) = K^{-1}$ has at most one nonzero root $\lambda$, and only if $K > K_l$.  This root satisfies $ -\sqrt{(KR)^2-1} \le  \lambda < 0$.  In addition, $\lambda = 0$ is a root  if and only if $K = K_c$. The equation $h_s(\lambda) = K^{-1}$ has $\lambda = 0$ as its only root in all cases.   Furthermore, the roots of the characteristic equations are in fact eigenvalues of $L$ on $E_c$ and $E_s$ respectively.

This completes the description of $\sigma_p(L)$.  Notice that we have to be a little careful here:  a root $\lambda \in \sigma(M)$ of one of  the characteristic equations is not automatically an eigenvalue of $L$, since the associated eigenvector $\eta$ might not be integrable w.r.t.~$g(\omega)$.  Fortunately as we shall see in Section~8, this doesn't happen.

Now what about the residual spectrum of $L$?  To answer this, we'll need a concrete description of the elements $\nu \in L^1(\Omega, W)^\ast$.  Let $W_0 = L^2(S^1, {\Bbb C}) \cap W$ be the subspace of $W$ consisting of measures of the form
$\phi \, dm$, where $\phi$ is $L^2$ on $S^1$.  The subspace $W_0$ naturally has the structure of a separable Hilbert space, and its image in $W$ is dense, which implies that  the natural map $L^1(\Omega, W)^\ast \to L^1(\Omega, W_0)^\ast$ is injective.  Now $L^1(\Omega, W_0)^\ast \cong L^\infty(\Omega, W_0^\ast)$ (see Lang [1993, p.~188] for this result), and the Hilbert space $W_0^\ast$ can be identified with the space of $L^2$ functions on $S^1$ with integral $0$.  So an element $\nu \in E_c^*$ can be represented as a function $\omega \mapsto \nu_\omega$, where $\nu_\omega$ is a function on $S^1$ for each frequency $\omega$.  If $\eta \in E_c$ the pairing $\langle \langle \eta, \nu\rangle \rangle$ of  $\eta$ and $\nu$ is given by integrating the function $\omega \mapsto \langle \nu_\omega, \eta_\omega \rangle $:
$$
\langle \langle \eta, \nu \rangle \rangle  = \int_\Omega \langle \nu_\omega, \eta_\omega \rangle g(\omega) d\omega.
$$
Now suppose  the operator $\lambda I - L$ does not have dense image on $E_c$.  Then there must be a nonzero element $\nu \in E_c^*$ such that
$$
 \langle \langle (\lambda I - L) \eta, \nu \rangle \rangle = \langle \langle (\lambda I - M) \eta , \nu \rangle \rangle - K (C\eta) \langle  \langle  D  \bigl(  (\sin  \theta)
 \rho \bigr),\nu \rangle\rangle  = 0
$$
for all $\eta \in E_c$ in the domain of $M$.  Since $M$ has only continuous spectrum, we cannot have
$\langle  \langle D  \bigl(  (\sin  \theta)  \rho \bigr),\nu  \rangle\rangle = 0$, so we may assume that $  \langle \langle D  \bigl(  (\sin  \theta)  \rho \bigr),\nu \rangle  \rangle=  K^{-1}$ if we like.  Then we have
 $$
 \langle\langle (\lambda I - M) \eta , \nu  \rangle\rangle =  C\eta
 $$
  for all $\eta \in E_c$ in the domain of $M$, or equivalently,
 $$
\langle \langle \eta , \nu \rangle \rangle =  C  (\lambda I - M) ^{-1} \eta
 $$
 for all $\eta \in E_c$ in the range of $\lambda I -M$, which is dense in $E_c$. The operator $(\lambda I - M_\omega) ^{-1}$ exists and is bounded for almost all $\omega$, and so we have
 $$\openup\jot\eqalign
 {
 \int_\Omega \langle \nu_\omega (\theta), \eta_\omega(\theta) \rangle g(\omega) \, d\omega &=  \int_\Omega \langle \cos \theta,  (\lambda I - M_\omega) ^{-1}\eta_\omega(\theta) \rangle g(\omega) \, d\omega \cr &=  \int_\Omega \langle  (\lambda I - M_\omega^\ast) ^{-1} \cos \theta, \eta_\omega(\theta) \rangle g(\omega) \, d\omega \cr
 }
 $$
 for all $\eta \in E_c$ in the range of $\lambda I -M$, where $M_\omega^\ast$ denotes the adjoint of the operator $M_\omega$.  This uniquely determines $\nu$:
 $$
\nu_\omega(\theta)  = (\lambda I - M_\omega^\ast) ^{-1} \cos \theta
$$
for almost all $\omega$.
 Now
$$\openup\jot\eqalign
 {
 \langle \langle D  \bigl(  (\sin  \theta)  \rho \bigr),\nu  \rangle \rangle &=  \int_\Omega \langle   (\lambda I - M_\omega^\ast) ^{-1} \cos \theta , D  \bigl(  (\sin  \theta) 
 \rho_\omega \bigr)\rangle g(\omega) d\omega  \cr
 &= \int_\Omega \langle \cos \theta, (\lambda I - M_\omega)^{-1} D  \bigl(  (\sin  \theta) 
 \rho_\omega \bigr) \rangle g(\omega) \, d\omega = h_c(\lambda), \cr
 }
 $$
so the self-consistency equation  $\langle\langle D  \bigl(  (\sin  \theta)  \rho \bigr),\nu  \rangle\rangle=  K^{-1}$ is exactly the same as before:  $h_c(\lambda) = K^{-1}$.  The same argument applies to $L$ on $E_s$.  But by Proposition 4, any roots of the characteristic equations are eigenvalues of $L$, so the only way $\lambda I - L$ can fail to have dense image is if $\lambda \in \sigma_p(L)$, and thus we conclude that $\sigma_r(L) = \emptyset$.
 \qed

We now turn to the proof of Proposition 3.
 \m
\noindent {\bf Proof of Proposition 3.}   We'll analyze the spectrum of $M$ on the spaces $E^l$ and $E^d$ separately and then combine the results.  The most general tangent vector $\eta \in E^l$ has the form
$$
\eta_\omega =  c (\omega)D\delta_{\theta_\omega}
$$
where the coefficient $c$ is an $L^1$-function of $\omega \in \Omega^l$ w.r.t.~the measure $g(\omega) d\omega$.  Let $\phi$ be any smooth function on $S^1$.  Then
$$
\langle \phi, M_\omega \eta_\omega \rangle = \langle \phi, -D(v_\omega \eta_\omega) \rangle = \langle v_\omega \phi', \eta_\omega \rangle = -c(\omega) (v_\omega \phi ' )' (\theta_\omega).
$$
But $v_\omega(\theta_\omega) = 0$, so
$$
\langle \phi, M_\omega \eta_\omega \rangle = -c(\omega) v_\omega'(\theta_\omega) \phi'(\theta_\omega) =c(\omega) KR \cos\theta_\omega \phi'(\theta_\omega),
$$
and hence
$$
M_\omega \eta_\omega = -KR c(\omega) \cos \theta_\omega D\delta_{\theta_\omega} .
$$
In other words, $M$ is just multiplication by the function $-KR \cos \theta_\omega$.  This explicit description shows that $M$ is a bounded operator on $E^l$, since the function $ \cos \theta_\omega$ is bounded.
$ (\lambda I -M_\omega)$ is multiplication by $(\lambda + KR \cos \theta_\omega)$ which is nonzero a.e.; hence $\lambda I - M$ has trivial kernel on $E^l$.

The inverse of $\lambda I - M$ on $E^l$ is multiplication by
 $ (\lambda + KR \cos \theta_\omega)^{-1}$;
this  operator  is bounded on $E^l \cong L^1 (\Omega^l, {\Bbb C})$   if and only if the function $\omega \mapsto (\lambda + KR \cos \theta_\omega)^{-1} $  is essentially bounded.  This is equivalent to $\lambda$ not in the image of the continuous function $\omega \mapsto - KR \cos \theta_\omega$ as $\omega$ ranges over the locked oscillators.  Hence the spectrum of $M$ on $E^l$ is $[-KR, -\sqrt{(KR)^2-1}]$ if $\rho$ is fully locked, and $[-KR, 0 ] $ if $\rho$ is partially locked.  For these $\lambda$ the operator $ (\lambda I -M)^{-1}$ is unbounded, but is defined on coefficient functions $c$ that are supported away from the roots of $\lambda + KR \cos \theta_\omega$, and hence is densely defined.  Therefore the spectrum of $M$ on $E^l$ is all continuous.

When the state $\rho$ is partially locked, we will show that the spectrum of $M$ on $E^d$ is ${\Bbb R}i$ and is all continuous, and this will complete the proof. For $\omega \in \Omega^d$, the operator $M_\omega$ is  the adjoint of the first-order differential
operator $N_\omega$ on $S^1$ defined by $N_\omega \phi = v_\omega \phi'$.  Note that $N_\omega$ is a closed, unbounded operator on the space of complex-valued functions $C^k(S^1, {\Bbb C})$, and is a bounded operator from $C^k(S^1,{\Bbb C})$ to $C^{k-1}(S^1, {\Bbb C})$.  We will see that for each $\omega$ with $|\omega| > KR$,  the spectrum $\sigma ( N_\omega) = \sigma_p(N_\omega) = \{ 2\pi i n C_\omega \, | \, n \in \Bbb Z \}$, where $C_\omega$ is the coefficient given by \Cdef.  So  if ${\rm Re}\, \lambda \ne 0$, then $(\lambda I - N_\omega)^{-1}$ exists and is a bounded (in fact compact) operator
on $C^k(S^1, {\Bbb C})$.  Therefore $ (\lambda I -M_\omega)^{-1} = \left((\lambda I - N_\omega)^{-1}\right)^\ast$ is a bounded operator on $C^k(S^1, {\Bbb C})^\ast$, with image in $ C^{k-1}(S^1, {\Bbb C})^\ast$.  Notice also that $(\lambda I - N_\omega) 1 = \lambda$, so $(\lambda I - N_\omega)^{-1} 1 = \lambda^{-1}$.  Hence if $\eta_\omega \in C^k(S^1, {\Bbb C})^\ast$ satisfies $\langle 1, \eta_\omega \rangle = 0$, then
$$
\langle 1, (\lambda I - M_\omega)^{-1}\ \eta_\omega \rangle = \langle (\lambda I - N_\omega)^{-1} 1, \eta_\omega \rangle = \lambda^{-1} \langle 1, \eta_\omega\rangle = 0.
$$
Therefore $ (\lambda I - M_\omega)^{-1}$ preserves the subspace $W \subset C^k(S^1, {\Bbb C})^\ast$, since $C^{k-1}(S^1, {\Bbb C})^\ast \subset C^k(S^1, {\Bbb C})^\ast_{abs}$.

The operator $N_\omega$ depends continuously on $\omega$, and hence so does the operator $(\lambda I - M_\omega)^{-1} $.  We will also show that the norm of $(\lambda I - N_\omega)^{-1}$ (as an operator mapping  $C^k(S^1, {\Bbb C})$ to itself) is bounded as a function of $\omega$. Since $\| \left((\lambda I - N_\omega)^{-1}\right)^\ast \| = \| (\lambda I - N_\omega)^{-1}\|$, all this proves that if $ {\rm Re} \,\lambda \ne 0$, then $\lambda I - M $ has a bounded inverse on $E^d$ defined for $\eta \in E^d$ by the rule
 $$
\left( (\lambda I - M)^{-1} \eta\right) _\omega =  \left((\lambda I - N_\omega)^{-1}\right)^\ast \eta_\omega.
$$

If $\lambda \in {\Bbb R}i$ but $\lambda \ne 0$, then $ (\lambda I - M_\omega)^{-1}$ is defined at all frequencies except $\pm \omega_n$, where
$$
\omega_n = \sqrt{(KR)^2 + {|\lambda|^2 \over n^2},} \quad n > 0
$$
is obtained by solving $\lambda = 2\pi i C_\omega n$ for $\omega_n$.  Note that the problem frequencies $\pm \omega_n$ have no limit point in the open intervals $\pm (KR, \infty)$.  $ (\lambda I - M)^{-1} \eta$ is defined for any $\eta \in E^d$ that vanishes on a neighborhood of $\{ \pm \omega_n \}$, and the set of such $\eta$ is dense in $E^d$.  This proves that $ (\lambda I - M)^{-1} $ is defined on a dense subspace of $E^d$; in other words, the image of $\lambda I - M$ is dense in $E^d$.  However, we claim that $\lambda I - M$ is not surjective, and hence $\lambda$ is in the continuous spectrum of $M$ on $E^d$.  To see this, let $\phi$ be a non-trivial solution to $(\lambda I - N_{\omega_n}) \phi = 0$ for some problem frequency $\omega_n$ ($\phi$ is guaranteed to be $C^\infty$).   Choose $\eta \in E^d$ such that $\langle \phi, \eta_\omega \rangle = 1$ for all $\omega$ in some neighborhood of $\omega_n$ (we can do this because $\phi$ is not constant), and suppose $(\lambda I - M)^{-1}\eta$ exists in $ E^d$. Then
$$
(\lambda I - N_\omega)\phi = \left( (\lambda I - N_{\omega_n}) + (N_{\omega_n} - N_\omega) \right) \phi =  (\omega_n - \omega) \phi ' ,
$$
and so for $\omega$ sufficiently close to $\omega_n$ we have
$$
1 = \langle \phi, \eta_\omega \rangle = \langle (\lambda I - N_\omega)\phi , (\lambda I  - M_\omega) ^{-1}\eta_\omega \rangle = (\omega_n - \omega) \langle \phi', (\lambda I  - M_\omega) ^{-1}\eta_\omega \rangle.
$$
Therefore
$$
1 \le| \omega - \omega_n| \, \|\phi'\| \|  (\lambda I  - M_\omega) ^{-1}\eta_\omega\|
$$
as $\omega \to \omega_n$, where the norms of $\phi'$ and $ (\lambda I  - M_\omega) ^{-1}\eta_\omega$ are taken in the spaces $C^k(S^1,{\Bbb C})$, $C^k(S^1, {\Bbb C})^\ast$ respectively.  But this implies that the function $\omega \mapsto  \|  (\lambda I  - M_\omega) ^{-1}\eta_\omega \| $ is not integrable, which is a contradiction.

Special care must be taken when $\lambda= 0$, since $0 \in \sigma_p(N_\omega)$ for all $\omega$. The spectrum of a closed operator is always a closed set, so $0$ is in the spectrum of $M$ on $E^d$.  $M$ has trivial kernel in $E^d$, since the only solutions to the equation $D(v_\omega \eta_\omega) = 0$ are multiples of $\rho_\omega$, but any $\eta \in E^d$ satisfies $\langle 1 , \eta_\omega \rangle = 0$ for  all $\omega$.   We can invert $M_\omega$ on the subspace $W \subset C^k(S^1, {\Bbb C})^\ast$ as follows:
$$
M_\omega ^{-1}  \eta _ \omega   =   - v_\omega^{-1} \left ( D^{-1} \eta_\omega  - { \langle v_\omega^{-1} ,D^{-1}\eta_\omega \rangle \over \langle v_\omega^{-1}, m \rangle}  m\right ),
$$
where $D^{-1} \eta_\omega$ denotes the unique distributional antiderivative of $\eta_\omega$ on $S^1$ determined by the requirement $\langle 1, D^{-1} \eta_\omega \rangle = 0$.
If $\eta_\omega = 0$ for $\omega$ in some neighborhood of $\pm KR$, then $M^{-1} \eta$ is perfectly well-behaved,  so $M^{-1}$ is defined on a dense subspace of $E^d$; in other words, the image of $M$ is dense.     Therefore we see that $M$ has empty point and residual spectra on $E^d$, and continuous spectrum ${\Bbb R}i$, as desired.

Now let's complete the analysis of the operator $N_\omega$.  Observe that 
$\lambda$ is an eigenvalue of $N_\omega$ if and only if the equation
$$
\lambda \phi - v_\omega \phi' = 0
$$
has a non-trivial solution on $S^1$.  This equation has general solution
$$
\phi (\theta) = k \exp \left (\lambda  \int _ 0 ^\theta v_\omega (\tau)^ {-1} d\tau \right),
$$
and assuming the constant $k \ne 0$ this function is periodic if and only if $\exp \bigl({ \lambda \over C_\omega} \bigr ) = 1$, which is equivalent to $\lambda = 2\pi i C_\omega n$ for some $n \in \Bbb Z$.  Therefore $\sigma_p (N_\omega) = 2 \pi i C_\omega \Bbb Z$.  If $\lambda \not \in 2 \pi i C_\omega \Bbb Z$ then we  rewrite the ODE
$$
\lambda \psi - v_\omega \psi' = \phi \cheqno\eqpsi
$$
as
$$
\psi' - \lambda v_\omega^{-1} \psi = - v_\omega^{-1} \phi.
$$
We can solve this equation with the aid of the integrating factor
$$
\gamma_\omega(\theta) = \exp \left( - \lambda \int_0^\theta v_\omega (\tau) ^{-1} d\tau \right);
$$
the unique periodic solution is
$$
\psi(\theta) = -\gamma_\omega(\theta) ^{-1}\left( \int_0^\theta \gamma_\omega (\tau)v_\omega ( \tau) ^{-1} \phi(\tau)  \, d\tau +(\gamma_\omega (2\pi) - 1 )^{-1} \int_0^{2 \pi} \gamma_\omega (\tau) v_\omega ( \tau)^{-1} \phi(\tau) \, d\tau \right).
$$
$\gamma_\omega(2 \pi) = \exp(-{\lambda \over C_\omega}) \ne 1$, so this function is well-defined.  The right-hand side defines the bounded operator $(\lambda I - N_\omega)^{-1}$ from $C^k(S^1, {\Bbb C})$ to $C^{k+1} (S^1, {\Bbb C})$, and so $\sigma ( N_\omega) = \sigma_p(N_\omega) = \{ 2\pi i n C_\omega \, | \, n \in \Bbb Z \}$.

Next we establish the necessary norm bounds for $(\lambda I - N_\omega)^{-1}$.  The differential equation \eqpsi\ has an irregular singular point at  the limiting values $\omega = \pm KR$, since the function $v_\omega = \omega - KR \sin \theta$ has a double root at $\pi \over 2$ when $\omega = KR$ (and a double root at$-{\pi \over 2}$ when $\omega = -KR$). So the fact that $\|(\lambda I - N_\omega)^{-1}\|$ is bounded as a function of $\omega$ is not trivial.  It suffices to prove the existence of constants $C_k$, $k \ge 0$, depending only on $\lambda$ and $KR$, such that
$$
\| (\lambda I - N_\omega)^{-1} \phi \|_k \le C_k \| \phi \|_k
$$
for any $\phi \in C^k(S^1, {\Bbb C})$, and any $\omega$ with $|\omega| > KR$.   For $k=0$, we take $\phi \in C(S^1,{\Bbb C})$ and put $\psi = (\lambda I - N_\omega)^{-1} \phi  \in C^1(S^1,{\Bbb C})$.  Suppose $|\psi(\theta)|^2$ has a positive maximum at $\theta = \theta_0$.
$$
{d \over d \theta} | \psi(\theta)|^2 = \psi '(\theta)\overline {\psi (\theta)} +\psi (\theta)\overline { \psi' (\theta) } = 2 {\rm Re}( \psi '(\theta)\overline { \psi (\theta) }),
$$
and so $ {\rm Re}( \psi '(\theta_0)\overline { \psi (\theta_0) } = 0$.  Multiply \eqpsi\ by $\overline {\psi(\theta_0)}$ and take real parts of both sides to obtain
$$
({\rm Re} \lambda )|\psi(\theta_0)|^2 = {\rm Re} ( \phi (\theta_0) \overline{\psi(\theta_0)});
$$
hence
$$
| \psi(\theta_0)| \le |{\rm Re} \lambda |^{-1} |\phi(\theta_0) |,
$$
and so we have $\| \psi \|_0 \le C_0 \| \phi \|_0$ with $C_0 = |{\rm Re} \lambda |^{-1}$.  This proves the base case $k = 0$ of our assertion.

Now we proceed by induction, and assume the existence of the constants $C_0, C_1, \dots , C_k$ has been established.   As before, let  $\psi = (\lambda I - N_\omega)^{-1} \phi$, where now $\phi \in C^{k+1}(S^1,{\Bbb C})$ and hence $\psi \in C^{k+2}(S^1,{\Bbb C})$.  Differentiating \eqpsi\ $k$ times gives
$$
v_\omega \psi^{[k+1]} =  \lambda \psi^{[k]}-\phi^{[k]} -  \sum_{j=0}^{k-1} {k \choose j} v_\omega^{[k-j]} \psi ^{[j+1]}.
$$
The terms $v_\omega^{[k-j]}$ are all bounded by $KR$, so by induction the right-hand side is bounded by $C'\| \phi \|_k$, and hence  $C'\| \phi \|_{k+1}$, for some constant $C'$.  If $|\psi^{[k+1]}|$ takes its maximum at $\theta = \theta_0$, then we obtain
$$
\| \psi^{[k+1]}\|_0 \le C' | \omega - KR \sin  \theta_0|^{-1} \| \phi \|_{k+1}.
$$
Now we differentiate \eqpsi\  one more  time and imitate the proof for the case $k=0$.  We write the result in the form
$$
(\lambda - (k+1)v_\omega') \psi^{[k+1]} - v_\omega \psi^{[k+2]} = \phi^{[k+1]} + \sum_{j=0}^{k-1} {k +1\choose j} v_\omega^{[k+1-j]} \psi ^{[j+1]},
$$
and observe that the right-hand side is bounded by $C'' \| \phi \|_{k+1}$ for some constant $C''$.  Multiply by $\overline {\psi^{[k+1]}(\theta_0)}$ and take real parts of both sides to obtain
$$
\| \psi^{[k+1]}\| _0 \le C''  | {\rm Re} \lambda + (k+1) KR \cos \theta_0|^{-1} \| \phi \|_{k+1}.
$$
Fortunately, the function
$$
m(\theta,  \omega) = \min \left (C' | \omega - KR \sin  \theta|^{-1} , C'' | {\rm Re} \lambda + (k+1) KR \cos \theta|^{-1} \right )
$$
is bounded for $\theta \in S^1$, $|\omega| > KR$.  (If not, we could construct a sequence $(\omega_n, \theta_n)$ such that $m(\omega_n, \theta_n) \to \infty$, but this implies that both
$$
|\sin \theta_n | \to 1 \quad {\rm and} \quad \cos \theta_n \to -{{\rm Re} \lambda \over (k+1) KR} \ne 0,
$$
which is impossible.) Therefore we have $||\psi^{[k+1]}\|_0 \le C''' \| \phi \|_{k+1}$, where $C'''$ is any bound on
$m(\theta, \omega)$, so we can take $C_{k+1} = C'''+C_k$, and we're done.

\qed

\noindent {\bf Remark on $\bf \sigma(L)$ for non-special positive states.}  In this case, we can carry out a similar analysis of the continuous spectrum $\sigma_c(M)$, but now it will contain positive real values, coming from the essential range of the function $\omega \mapsto -KR \cos \theta_\omega^* = KR \cos \theta_\omega$.  The same reasoning used in our main theorem shows that these values are also contained in $\sigma(L)$, so these states are not linearly stable.  We omit the details since these states are of minor importance to us.

\b
\b
\noindent {\bf 7.  Characteristic Functions for $\bf L$}
\b

Our next task is to compute the characteristic functions $h_c$ and $h_s$, given by
$$
\eqalign
{
 h_c(\lambda) &= \int _\Omega \langle \cos \theta, (\lambda I - M_\omega)^{-1} D  \bigl(  (\sin  \theta)
 \rho_\omega \bigr) \rangle g(\omega) d\omega, \cr
  h_s(\lambda) &=- \int _\Omega \langle \sin \theta, (\lambda I - M_\omega)^{-1} D  \bigl(  (\cos  \theta)
 \rho_\omega \bigr) \rangle g(\omega) d\omega.
}
$$
To do this, we'll have to compute $ (\lambda I - M_\omega)^{-1} D  \bigl(  (\sin  \theta)\bigr)
$ and $ (\lambda I - M_\omega)^{-1} D  \bigl(  (\cos  \theta) $ explicitly.  Let's begin with the locked frequencies $\omega \in \Omega^l$.  As we saw in the previous section, for these frequencies $(\lambda I - M_\omega)^{-1} $ is just multiplication by $(\lambda +KR \cos \theta_\omega)^{-1}$.  Hence we obtain the contributions $h^l_c$ and $h^l_s$ resp.~from the locked oscillators to the characteristic functions $h_c$ and $h_s$:
$$\openup\jot\eqalign
{
h_c^l(\lambda) &=  \int_{\Omega^l} (\lambda +KR \cos \theta_\omega)^{-1}  \langle \cos \theta,   D  \bigl(  (\sin  \theta)
 \rho_\omega \bigr) \rangle \, g(\omega) d\omega  \cr
&=  \int_{\Omega^l} (\lambda +KR \cos \theta_\omega)^{-1}  \langle \sin \theta,    (\sin  \theta)
 \rho_\omega \bigr) \rangle \, g(\omega) d\omega  \cr&=  \int_{\Omega^l} { \sin^2 \theta_\omega   \over \lambda  +KR \cos  \theta_\omega}   \, g(\omega) d\omega \cr
&= {1 \over (KR)^2} \int_{-KR}^{KR} {\omega^2 \over  \lambda +  \sqrt{  (KR)^2 -\omega ^2} }  \, g(\omega)  d \omega \cr
} \cheqno\eqhcl
$$
and
$$\openup\jot\eqalign
{
h_s^l(\lambda) &=  -\int_{\Omega^l} (\lambda +KR \cos \theta_\omega)^{-1} \langle \sin \theta, D  \bigl(  (\cos  \theta)
 \rho_\omega \bigr)   \rangle \, g(\omega) d\omega  \cr
 &= \int_{\Omega^l} { \cos^2 \theta_\omega    \over \lambda  +KR \cos  \theta_\omega}  \, g(\omega)  d\omega \cr
&=  {1 \over (KR)^2} \int_{-KR}^{KR} {(KR)^2 -\omega^2 \over  \lambda +  \sqrt{  (KR)^2 -\omega ^2} } \, g(\omega)  d \omega .\cr
}
\cheqno\eqhsl
$$
The functions $h_c^l(\lambda)$, $h_s^l(\lambda)$ are (up to the constant $K$) the $N \to \infty$ limits of the rational functions $R_s(\lambda)$, $R_c(\lambda)$ resp.~that we defined in our study of the finite-$N$ Kuramoto model [Mirollo and Strogatz 2005], as was  to be expected.  After all, completely locked states are just the infinite-$N$ analogues of  fixed points for the finite-$N$ Kuramoto model.  However the partially locked states have no finite-$N$ analogues, so we should expect to see something new there.

Now suppose $\rho$ is partially locked, and consider the drifting frequencies $\omega \in \Omega^d$.
For almost all $\omega \in \Omega^d$, the distribution $ (\lambda I - M_\omega)^{-1} D  \bigl(  (\sin  \theta)  \rho_\omega \bigr)$ will be a smooth measure of the form $\alpha_\omega (\theta) d\theta$,
where $\alpha_\omega$ is a smooth function on $S^1$.  The operator $\lambda I - M_\omega$ applied to the
measure $\alpha_\omega (\theta) d\theta$ gives the measure $\left (\lambda \alpha_\omega (\theta)+
(v_\omega \alpha_\omega)' (\theta) \right) d\theta$,
and $D  \bigl(  (\sin  \theta)  \rho_\omega \bigr)$ is just the measure $C_\omega \left (\sin \theta \over v_\omega(\theta) \right)'d\theta$.
Therefore $\alpha_\omega$ must satisfy  the first-order ODE
$$
\lambda \alpha_\omega (\theta)  +\left (v_\omega (\theta) \alpha_\omega (\theta)\right)' = C_\omega\left ({\sin \theta \over v_\omega(\theta)}\right )' .
$$
Our strategy for solving this equation is to express $\alpha_\omega$ in the form $\beta_\omega / v_\omega^2$, so as to clear out the denominator $v_\omega^2$ above. The corresponding equation  for $\beta_\omega$ is
$$
\lambda \beta_\omega(\theta) + v_\omega(\theta)  \beta_\omega'(\theta) - v_\omega' (\theta)\beta_\omega(\theta) = C_\omega (  v_\omega(\theta)\cos\theta -  v_\omega' (\theta) \sin\theta). \cheqno\eqbeta
$$
Now a fortunate miracle occurs.  The functions $\sin \theta$, $\cos \theta$ and $v_\omega(\theta)$ are all contained in the vector space $V$ spanned by $1$, $\cos \theta$ and $\sin \theta$.  Notice that $V$ is closed under the operation
$$
\{ \phi, \psi \} = \phi \psi' - \phi' \psi,
$$
so we restrict our search for solutions $\beta_\omega$ to \eqbeta\ to functions of the form
$$
\beta_\omega(\theta) = c_0(\omega) +c_1(\omega)  \cos \theta + c_2 (\omega) \sin \theta,
$$
with coefficients $c_i$ that depend on $\omega$.  (The miracle is that the three-dimensional subspace consisting of measures $\rho_\omega$ of the form $(\beta_\omega / v_\omega^2) d\theta$, with $\beta_\omega$ being a linear combination of $1$, $\sin \theta$ and $\cos \theta$, is invariant under the differential operator $M_\omega$.)  Substituting $v_\omega(\theta) = \omega -KR \sin\theta$  in \eqbeta\ gives
$$
\openup\jot\eqalign
{
&\left (\lambda c_0 (\omega) + KR c_1(\omega)\right) + \left ( KRc_0(\omega)+\lambda c_1(\omega) +\omega c_2(\omega)\right) \cos \theta +\left (-\omega c_1(\omega)+\lambda c_2(\omega)\right ) \sin \theta \cr
& = \omega C_\omega  \cos \theta. \cr
}
$$
We equate coefficients and solve for $c_i(\omega) $ to obtain
$$
\openup\jot\eqalign
{
c_0 (\omega)&=- { KR \omega C_\omega  \over \lambda^2+\omega^2 -(KR)^2} ,\cr
c_1 (\omega)&= {\lambda \omega C_\omega \over \lambda^2+\omega^2 -(KR)^2} ,\cr
c_2 (\omega) &= { \omega ^2 C_\omega \over  \lambda^2+\omega^2 -(KR)^2} .\cr
}
$$
So the contribution  $h^d_c$ to $h_c$ from the drifting frequencies is given by
$$
\openup\jot\eqalign
{
h_c^d(\lambda) &= \int _{\Omega^d} \langle \cos \theta, (\lambda I - M_\omega)^{-1} D  \bigl(  (\sin  \theta)
 \rho_\omega \bigr) \rangle g(\omega) d\omega
\cr
&=  \int\limits_{|\omega| \ge KR} \left (\int_0^{2 \pi} \cos\theta (c_0(\omega) +c_1 (\omega)\cos \theta + c_2 (\omega)  \sin \theta) v_\omega^{-2}(\theta) \, d\theta\right)  g(\omega) d\omega \cr
&= \int\limits_{|\omega| \ge KR}c_1(\omega) \left ( \int_0^{2 \pi} { \cos^2\theta  \over (\omega -KR\sin\theta)^2} \, d\theta  \right)  g(\omega) d\omega \cr
}
$$
(the other two integrals vanish because the integrands  have periodic antiderivatives on $S^1$).  We can evaluate the inner integral using integration by parts:
$$
\openup\jot\eqalign
{
 \int_0^{2 \pi} { \cos\theta  \over (\omega -KR\sin\theta)^2}\cos\theta \, d\theta
 &= {1 \over KR} \int_0^{2 \pi} {\sin\theta \over \omega-KR \sin \theta} \,d\theta\cr
 &={1 \over (KR)^2} \int_0^{2 \pi} \left( {\omega \over \omega - KR \sin\theta} - 1\right) d\theta \cr
 &= {1 \over (KR)^2}( \omega C_\omega^{-1}-2\pi). \cr
}
$$
Therefore
$$
\openup\jot\eqalign
{
h_c^d(\lambda)&= {1 \over (KR)^2} \int\limits_{|\omega| \ge KR} { \lambda \omega  C_\omega \over \lambda^2+\omega^2 -(KR)^2 } (\omega C_\omega^{-1}-2\pi) g(\omega)  d\omega \cr
&= {1 \over (KR)^2} \int\limits_{|\omega| \ge KR} { \lambda  \over \lambda^2+\omega^2 -(KR)^2 } ( \omega^2- 2 \pi  \omega C_\omega) g(\omega)  d\omega \cr
&=  \int\limits_{|\omega| \ge KR} { \lambda \over \lambda^2+\omega^2 -(KR)^2 } \cdot {|\omega| \over |\omega| +\sqrt{ \omega^2 - (KR)^2 }} \,  g(\omega)  d\omega. \cr
}
\cheqno\eqhcd
$$

We repeat this procedure to compute the smooth measures $(\lambda I - M_\omega)^{-1} D  \bigl(  (\cos  \theta)  \rho_\omega \bigr)$ for $\omega \in \Omega^d$; express this measure in the form $\beta_\omega(\theta) d\theta$,
where now $\beta_\omega$ satisfies the equation
$$
\lambda \beta_\omega(\theta) + v_\omega(\theta)  \beta_\omega'(\theta) - v_\omega' (\theta)\beta_\omega(\theta) = -C_\omega (  v_\omega(\theta)\sin\theta +  v_\omega' (\theta) \cos \theta).
$$
We put $\beta_\omega(\theta) = c_0(\omega) +c_1(\omega)  \cos \theta + c_2 (\omega) \sin \theta
$
and solve for the coefficients $c_i$ as before, this time obtaining
$$
\openup\jot\eqalign
{
c_0 (\omega)&= { KR \lambda  C_\omega  \over \lambda^2+\omega^2 -(KR)^2} ,\cr
c_1 (\omega)&= { C_\omega (\omega^2 - (KR)^2) \over \lambda^2+\omega^2 -(KR)^2} ,\cr
c_2 (\omega) &=- {\lambda  \omega  C_\omega \over  \lambda^2+\omega^2 -(KR)^2} .\cr
}
$$
The contribution  $h^d_s$ to $h_s$ from the drifting frequencies is given by
$$
\openup\jot\eqalign
{
h_s^d(\lambda) &=-  \int _{\Omega^d} \langle \sin \theta, (\lambda I - M_\omega)^{-1} D  \bigl(  (\cos  \theta)
 \rho_\omega \bigr) \rangle g(\omega) d\omega
\cr
&= - \int\limits_{|\omega| \ge KR} \left (\int_0^{2 \pi} \sin\theta (c_0(\omega) +c_1 (\omega)\cos \theta + c_2 (\omega)  \sin \theta) v_\omega^{-2}(\theta) \, d\theta\right)  g(\omega) d\omega \cr
&= -\int\limits_{|\omega| \ge KR}  \int_0^{2 \pi} {c_0 (\omega)  \sin \theta + c_2(\omega) \sin^2 \theta \over (\omega -KR\sin\theta)^2} \, d\theta.  \cr
}
$$
Substituting the coefficients $c_0$ and $c_2$ gives
$$
\openup\jot\eqalign
{
h_s^d(\lambda) &= \int\limits_{|\omega| \ge KR}{ \lambda C_\omega \over \lambda^2+\omega^2 -(KR)^2 }\left (  \int_0^{2 \pi} { -KR \sin \theta +\omega \sin^2 \theta \over (\omega -KR\sin\theta)^2} \, d\theta  
\right)  g(\omega) d\omega \cr
&= \int\limits_{|\omega| \ge KR}{ \lambda C_\omega \over \lambda^2+\omega^2 -(KR)^2 }\left (  \int_0^{2 \pi}  \left( {1 \over \omega -KR \sin \theta} -{ \omega \cos^2 \theta \over (\omega -KR\sin\theta)^2}  \right)\, d\theta   \right)  g(\omega) d\omega \cr
&=  \int\limits_{|\omega| \ge KR} { \lambda \over \lambda^2+\omega^2 -(KR)^2 } \cdot { \sqrt{ \omega^2 - (KR)^2 } \over |\omega| +\sqrt{ \omega^2 - (KR)^2 }}  \,g(\omega) d\omega. \cr
}
\cheqno\eqhsd
$$
So the characteristic functions $h_c$ and $h_s$ are given by $h_c(\lambda) = h_c^l(\lambda) + h_c^d (\lambda)$ and $h_s(\lambda) = h_s^l(\lambda) + h_s^d (\lambda)$, with the functions $h_c^l$, $h_c^d$, $h_s^l$ and $h_s^d$ given by \eqhcl , \eqhsl , \eqhcd\ and \eqhsd\ above.  This completes the derivation of the characteristic functions.

\b
\b
\b\noindent{\bf  8. Roots of the Characteristic Equations}
\b

Our final task is to prove Proposition 4, which we restate here for convenience:

  \proclaim Proposition 4.  The equation $h_c(\lambda) = K^{-1}$ has at most one  nonzero root $\lambda$, and only if $K > K_l$.  This root satisfies $ -\sqrt{(KR)^2-1} \le  \lambda < 0$.  In addition, $\lambda = 0$ is a root  if and only if $K = K_c$. The equation $h_s(\lambda) = K^{-1}$ has $\lambda = 0$ as its only root in all cases.   Furthermore, the roots of the characteristic equations are in fact eigenvalues of $L$ on $E_c$ and $E_s$ respectively.

 \pf Notice first that
 $$
 h_s^l(0) =  {1 \over (KR)^2} \int_{-KR} ^{KR} \sqrt{  (KR)^2 -\omega ^2}  \, g(\omega)  d \omega ={KR^2 \over (KR)^2} = K^{-1}
 $$
 and $h_s^d(0) = 0$ in the partially locked case, so $\lambda = 0$ is always a root of the characteristic equation $h_s(\lambda) = K^{-1}$.  This is no surprise, since the rotational symmetry of the Kuramoto model implies that $\lambda = 0$ is always an eigenvalue of $L$.  The corresponding eigenvector is just $D\rho \in E_s$: we have $S(D\rho) = -R$, and so
 $$
 \openup\jot\eqalign
{
 L(D\rho)_\omega &= -D(v_\omega D\rho_\omega -KR (\cos \theta) \rho_\omega) \cr
 &= -D^2(v_\omega \rho_\omega) = 0.
 }
 $$

Next, let's consider the fully locked case, so $\Omega = [-1,1]$,  $K \ge K_l$ (equivalently, $KR \ge 1$), $h_c = h_c^l$ and $h_s = h_s^l$.
If $\lambda \in \Bbb C$ has nonzero imaginary part, then the same is true for
$h_c(\lambda)$ and $h_s(\lambda)$; to see this, multiply the numerator and denominator in the integrands for $h_c^l(\lambda)$ and $h_s^l(\lambda)$  by $\overline \lambda +\sqrt{(KR)^2 - \omega^2}$. Hence the characteristic equations can have only real roots in this case.
So let $\lambda \in \Bbb R$.
If $\lambda \le -KR$ then $h_c(\lambda)$, $h_s(\lambda) < 0$, so $\lambda$ is not a root of the characteristic equations.   (Note that $h_c(-KR)$ is well-defined, whereas $h_s(-KR) = -\infty$.)
$h_c(\lambda)$ and $h_s(\lambda)$ are undefined for $-KR < \lambda <  -\sqrt{(KR)^2-1}$,
since the integrands in the formulas for $h_c(\lambda)$ and $h_s(\lambda)$ have simple poles at $\omega = \pm \sqrt{(KR)^2 - \lambda^2} \in (-1,1)$.  The functions $h_c$ and $h_s$ are both defined and positive on $(  -\sqrt{(KR)^2-1}, \infty)$, but might take the value $+\infty$ at $\lambda =  -\sqrt{(KR)^2-1}$.  And both functions are strictly decreasing on the interval $[  -\sqrt{(KR)^2-1}, \infty)$, so the characteristic equations each can have at most one root here.  Since we already saw that $h_s(0) = K^{-1}$ in all cases, we conclude that there are no other roots to the characteristic equation $h_s(\lambda) = K^{-1}$ in the fully locked case.

 Next we claim that $h_c(0) \le K^{-1}$, which implies that any root $\lambda$  of the equation  $h_c(\lambda) = K^{-1}$ must satisfy $\lambda \le 0$.  To see this, observe that
$$
 \openup\jot\eqalign
{
 h_c(0) - K^{-1} &= {1 \over (KR)^2} \int_{-KR}^{KR}\left(   {\omega^2 \over   \sqrt{  (KR)^2 -\omega ^2} }  -\sqrt{  (KR)^2 -\omega ^2}  \right) g(\omega)  d \omega \cr
&=  2\int_0^1 \left (    {s^2 \over   \sqrt{  1 -s^2} } - \sqrt{ 1 -s^2}  \right )  g(KRs) ds.
}
$$
The function $h(s) =  {s^2 \over   \sqrt{  1 -s^2} }  -  \sqrt{ 1 -s^2}  $ changes sign from negative to positive at $s_0 = {\sqrt 2 \over 2}$, and $g$ is non-increasing on $[0,1]$ by assumption.  Therefore
$$
 \openup\jot\eqalign
{
{1 \over 2} \left( h_c(0)-K^{-1}   \right) &=  \int_0^ {s_0} h(s)  g(KRs) ds + \int_ {s_0}^1h(s)  g(KRs) ds \cr
& \le g(KR s_0)  \int_0^  {s_0} h(s) ds + g(KR s_0)  \int_ {s_0}^1 h(s)  ds \cr
&= g(KRs_0)  \int_0^1 \left (  {s^2 \over   \sqrt{  1 -s^2} }  -  \sqrt{ 1 -s^2}  \right )   ds = 0.
}
$$
This inequality is strict except in one special case: when $KR = 1$  and $g$ is constant on $[0,1]$; recall that in this case the two critical coupling constants $K_c$ and $K_l$ are equal.  So in the fully locked case with $K > K_l$ the equation $h_c(\lambda) = K^{-1}$ has at most one root, which satisfies $-\sqrt{(KR)^2 -1} < \lambda < 0$.  If $K = K_l$ then the equation $h_c(\lambda) = K^{-1}$ has no roots, except for $\lambda = 0$ in the special case where $g$ is constant on $[0,1]$, or equivalently when $K =K_l = K_c$ .

If $K > K_l$, then the characteristic equation $h_c(\lambda) = K^{-1}$ has a root in the interval $[  -\sqrt{(KR)^2-1}, 0)$ if and only if $h_c(-\sqrt{(KR)^2-1}) \ge K^{-1}$.
 If we examine the formula for $h_c(-\sqrt{(KR)^2-1})$, we see that we can make this value as small or as large as we like by varying the density function (to make $h_c(-\sqrt{(KR)^2-1})$ small, take the density function to be concentrated at $\omega = 0$, where the integrand vanishes).  So it is possible that $h_c$ has no roots in $[  -\sqrt{(KR)^2-1}, 0)$.  If the endpoint $\lambda = -\sqrt{(KR)^2-1} \in \sigma(M)$ happens to be a root, then we need to check that the associated eigenvector  $\eta$ given by
 $$
 \eta_\omega ={\omega \over KR}  (\lambda + \sqrt{(KR)^2 - \omega^2} )^{-1} D \delta_{\theta_\omega}
 $$
 is in fact an element of $E_c$.  The coefficient function above becomes infinite  as $\omega \to \pm 1$, but is nevertheless integrable against $g(\omega)$, since the integrand in the formula for $h_c(-\sqrt{(KR)^2-1})$ has the same asymptotic behavior as $\omega \to \pm 1$.   So $\eta$ is a bona fide element of $E_c$, and $\lambda$ is an eigenvalue of $L$ on $E_c$.  The same thing happens in the special case with $KR = 1$ and $g$ constant on $[-1,1]$: we have
 $$
 \eta_\omega ={\omega \over \sqrt{1 - \omega^2} }D \delta_{\theta_\omega}
$$
 and the coefficient function is integrable on $[-1,1]$, so $\lambda = 0$ is an eigenvalue of $L$ on $E_c$. This completes the proof for the fully locked case.

Now we turn to the partially locked case $K < K_l$, and prove that both characteristic equations have no roots $\lambda \ne 0$.  Let's begin with $\lambda \in \Bbb R$.  If $\lambda \le -KR$, then $h_c(\lambda)$, $h_s(\lambda) < 0$, and if $-KR < \lambda <0$, then $h_c(\lambda)$ and $h_s(\lambda)$ are undefined.  To complete the proof for $\lambda \in \Bbb R$, we will establish that $h_c(\lambda) < K^{-1}$ and $h_s(\lambda) < K^{-1}$ for all $\lambda > 0$.   To simplify the notation a bit, replace $\lambda$ with $KR \lambda$;  then the  first inequality is equivalent to
$$
 \int_0^1  \left(  {s^2 \over \lambda + \sqrt{1-s^2}} -  \sqrt{  1 -s^2} \right) g(KRs)   ds  + \int_1^\infty {\lambda \over \lambda^2+s^2-1} \cdot {s \over s+\sqrt{s^2-1}} \, g(KRs) ds  < 0
$$
for all $\lambda  > 0$.   As before, the function $h(s) =  {s^2 \over \lambda + \sqrt{1-s^2}} -  \sqrt{  1 -s^2} $ changes sign from negative to positive at a unique point $s_ 0 \in (0,1)$,  so
$$
 \int_0^1  \left(  {s^2 \over \lambda + \sqrt{1-s^2}} -  \sqrt{  1 -s^2} \right) g(KRs)   ds \le g(KRs_0)
 \int_0^1  \left(  {s^2 \over \lambda + \sqrt{1-s^2}} -  \sqrt{  1 -s^2} \right) ds;
 $$
the integral on the right is negative for $\lambda > 0$ and $g$ is non-increasing, so we have
$$
 \int_0^1  \left(  {s^2 \over \lambda + \sqrt{1-s^2}} -  \sqrt{  1 -s^2} \right) g(KRs)   ds \le g(KR)
 \int_0^1  \left(  {s^2 \over \lambda + \sqrt{1-s^2}} -  \sqrt{  1 -s^2} \right) ds.
$$
For the second integral, observe that
$$
\int_1^\infty {\lambda \over \lambda^2+s^2-1} \cdot {s \over s+\sqrt{s^2-1}} \, g(KRs) ds  < g(KR)  \int_1^\infty {\lambda \over \lambda^2+s^2-1} \cdot {s \over s+\sqrt{s^2-1}} \, ds
$$
 (strict inequality holds because $g(KR) > 0$ and $g(\omega) \to 0$ as $\omega \to \infty$).  So it suffices to prove that
$$
 \int_0^1  \left(  {s^2 \over \lambda + \sqrt{1-s^2}} -  \sqrt{  1 -s^2} \right)  ds  + \int_1^\infty {\lambda \over \lambda^2+s^2-1} \cdot {s \over s+\sqrt{s^2-1}} \,  ds   \le 0
$$
for all $\lambda > 0$.  These integrals can be evaluated explicitly, and in fact equality holds above:
$$
 \int_0^1  {s^2 \over  \lambda + \sqrt{1-s^2} } \,  ds + \int_1^\infty {\lambda \over \lambda^2+s^2-1} \cdot {s \over s+\sqrt{s^2-1}} \, ds  = \int_0^1 \sqrt{1-s^2} \, ds  ={\pi \over 4}
 $$
if $\Re \lambda > 0$.  The relevant explicit formulas are
$$
 \int_0^1  {s^2 \over  \lambda + \sqrt{1-s^2} } \,  ds =  -{\pi \over 2} \lambda^2 + \lambda +{\pi \over 4}  + \lambda \sqrt{\lambda^2-1} \tan^{-1} \sqrt {\lambda^2-1}
 $$
 and
 $$
\int_1^\infty {\lambda  \over \lambda^2+s^2-1} \cdot {s \over s+\sqrt{s^2-1}} \, ds =   {\pi \over 2} \lambda^2  - \lambda - \lambda \sqrt{\lambda^2-1} \tan^{-1} \sqrt {\lambda^2-1}
 $$
for $\lambda > 1$.   Keep in mind that the integrals above are analytic functions on the domain $ \Re \lambda > 0$, so proving they agree for $\lambda > 1$ is sufficient.

The proof of the  inequality $h_s(KR \lambda) < K^{-1}$ is similar: we need
$$
\int_0^1  \left(  {1-s^2 \over \lambda + \sqrt{1-s^2}} -  \sqrt{  1 -s^2} \right) g(KRs)   ds +
\int_1^\infty {\lambda \over \lambda^2+s^2-1} \cdot {\sqrt{s^2 -1} \over s+\sqrt{s^2-1}} \, g(KRs) ds <  0$$
for all $\lambda > 0$.  We have
$$
 \openup\jot\eqalign
{
\int_0^1  \left(  {1-s^2 \over \lambda + \sqrt{1-s^2}} -  \sqrt{  1 -s^2} \right) g(KRs)   ds&=- \lambda  \int_0^1  {\sqrt{1-s^2} \over \lambda +  \sqrt{1-s^2} } \,   g(KRs) ds \cr
&\le -\lambda g(KR)  \int_0^1  {\sqrt{1-s^2} \over \lambda +  \sqrt{1-s^2} } \, ds
}
$$
and
$$
\int_1^\infty {\lambda \over \lambda^2+s^2-1} \cdot {\sqrt{s^2 -1} \over s+\sqrt{s^2-1}} \, g(KRs) ds <  \lambda g(KR) \int_1^\infty {1 \over \lambda^2+s^2-1} \cdot {\sqrt{s^2 -1} \over s+\sqrt{s^2-1}} \,ds
$$
as before.  This time we find that
$$
 \openup\jot\eqalign
{
\int_1^\infty {1 \over \lambda^2+s^2-1} \cdot {\sqrt{s^2 -1} \over s+\sqrt{s^2-1}} \, ds  &=   \int_0^1
{  \sqrt{  1 -s^2}  \over \lambda + \sqrt{1-s^2}} \, ds \cr
&= - {\pi \over 2} \lambda + 1 +{\lambda^2 \over  \sqrt{\lambda^2-1}}\tan^{-1} \sqrt {\lambda^2-1} \cr
 }
$$
for all $\lambda > 1$, and  this completes the proof that the characteristic equations have no non-zero real roots.

Now let's rule out any complex roots.  Since $h_c(\overline \lambda) = \overline {h_c(\lambda)}$ and $h_s(\overline \lambda) = \overline {h_s(\lambda)}$, it suffices to prove that $\Im \lambda > 0$ implies $\Im h_c(\lambda) < 0 $ and $\Im h_s(\lambda) < 0$.  Let $\Im \lambda > 0$.  If $\lambda$ is pure imaginary, then $h_c(\lambda)$ and $h_s(\lambda)$ are defined only if $\Omega = [-1,1]$ and $\Im \lambda \ge KR$, and it's easy to see that $\Im h_c(\lambda) < 0 $ and $\Im h_s(\lambda) < 0$ in this case (all the integrands have negative imaginary part).  So assume $\Re \lambda \ne 0$.
Replacing $\lambda$ with $KR \lambda$ as before, we see that it suffices to prove
 $$
 \Im \int_0^1  {s^2 \over \lambda + \sqrt{1-s^2}}\,  g(KRs) ds +  \Im \int_1^\infty  {\lambda \over \lambda^2+s^2-1}\cdot {s \over s+\sqrt{s^2-1}} \, g(KRs) ds
 <0
 $$
 and
 $$
  \Im \int_0^1  {1 - s^2 \over \lambda + \sqrt{1-s^2}}\,  g(KRs) ds +  \Im \int_1^\infty  {\lambda \over \lambda^2+s^2-1}\cdot {\sqrt{s^2-1} \over s+\sqrt{s^2-1}} \, g(KRs) ds
 <0
 $$
 for $\Im \lambda > 0$.
 The proofs of these two inequalities are identical, so we'll present only  the first one.
 $$
\Im \left({s^2 \over \lambda + \sqrt{1-s^2}} \right)< 0
$$
for $ 0< s < 1$, so
$$
 \Im \int_0^1  {s^2 \over \lambda + \sqrt{1-s^2}}\,  g(KRs) ds \le g(KR)  \Im\int_0^1 {s^2 \over \lambda + \sqrt{1-s^2}}\,  ds.
$$
The function
$$
\Im \left( {\lambda \over \lambda^2+s^2-1}\right) = \Im \lambda \left ({s^2 - |\lambda|^2 -1 \over |\lambda^2 +s^2-1|^2}  \right)
$$
changes sign from negative to positive at $s_0 = \sqrt{|\lambda|^2+1}$.  So by the same argument as before,
$$
 \Im \int_1^\infty  {\lambda \over \lambda^2+s^2-1}\cdot {s \over s+\sqrt{s^2-1}} \, g(KRs) ds  < g(KR s_0) \Im   \int_1^\infty  {\lambda \over \lambda^2+s^2-1}\cdot {s \over s+\sqrt{s^2-1}} \,  ds.
$$
Therefore
$$
 \openup\jot\eqalign
{
& \Im \int_0^1  {s^2 \over \lambda + \sqrt{1-s^2}}\,  g(KRs) ds +  \Im \int_1^\infty  {\lambda \over \lambda^2+s^2-1}\cdot {s \over s+\sqrt{s^2-1}} \, g(KRs) ds \cr
& < g(KR)  \Im \int_0^1  {s^2 \over \lambda + \sqrt{1-s^2}}\,  ds + g(KRs_0) \Im \int_1^\infty  {\lambda \over \lambda^2+s^2-1}\cdot {s \over s+\sqrt{s^2-1}} \, ds \cr
& \le g(KRs_0)  \Im \int_0^1  {s^2 \over \lambda + \sqrt{1-s^2}}\,  ds + g(KRs_0) \Im \int_1^\infty  {\lambda \over \lambda^2+s^2-1}\cdot {s \over s+\sqrt{s^2-1}} \, ds \cr
&=g(KRs_0) \Im \left ( \int_0^1  {s^2 \over \lambda + \sqrt{1-s^2}}\,  ds +  \int_1^\infty  {\lambda \over \lambda^2+s^2-1}\cdot {s \over s+\sqrt{s^2-1}} \, ds \right). \cr
}
$$
The right hand side is $0$ if $\Re \lambda > 0$, so we're done in this case.  If $\Re \lambda < 0$, let $\lambda^\ast = -\overline \lambda$ be the reflection of $\lambda$ in the imaginary axis.  Then
$$
\Im  \left({s^2 \over \lambda + \sqrt{1-s^2}} \right) < \Im  \left({s^2 \over \lambda^\ast + \sqrt{1-s^2}} \right)
$$ for $0 <s <1$ and
$$
\Im \left ({\lambda \over \lambda^2+s^2-1} \right )= \Im \left( {\lambda^\ast  \over( \lambda^\ast)^2+s^2-1} \right)
$$
for all $s \ge 1$, so we are done by the previous case.

To finish up, we need to see when $\lambda = 0$ is a root of the characteristic equation $h_c(\lambda) = K^{-1}$.  The same argument as in the fully locked case shows that this holds if and only if $g$ is constant on $[-KR, KR]$, which is equivalent to $K = K_c$.  Since $0 \in \sigma(M)$, we need to check that the associated eigenvector $\eta$ given by the formulas in Section~7 is a bona fide element of the tangent space $E$.  If $|\omega| \le KR$, $\eta$ is given by
$$
\eta_\omega = {\omega \over KR} \cdot {1 \over \sqrt{(KR)^2 - \omega^2}} D \delta_{\theta_\omega},
$$
and if $|\omega| > KR$, $\eta_\omega (\theta) = \alpha_\omega (\theta) d\theta$ with
$$
\alpha_\omega(\theta) = {1 \over 2\pi} \cdot {|\omega| \over \sqrt{\omega^2 - (KR)^2}} \cdot  {-KR +\omega \sin \theta \over( \omega - KR \sin \theta)^2 }  .
$$
The function $\omega \mapsto ((KR)^2 - \omega^2)^{-{1\over 2}}$ has an integrable singularity at $\omega = \pm KR$, and fortunately, the same thing happens for the drifting frequencies.   To see this, express $\omega = KR (1+\epsilon)$, with $0 < \epsilon < 1$,  to analyze the singularity at $\omega = KR$ (the singularity at $-KR$ has the same behavior).  Then
 $$
 \alpha_\omega(\theta) =  {1 \over 2\pi KR } \cdot {1+ \epsilon  \over \sqrt{2\epsilon+\epsilon^2}} \cdot {-1 + (1 + \epsilon) \sin \theta \over (1+\epsilon - \sin \theta)^2 }  = {1 \over 2\pi KR } \cdot {1+ \epsilon \over \sqrt{2\epsilon+\epsilon^2}}  \left( \log (1+ \epsilon - \sin\theta) \right)'' .
$$
If $\phi$ is any $C^\infty$ function on $S^1$, then
$$
\langle \phi , \eta_\omega \rangle =  {1 \over 2\pi KR} \cdot {1+\epsilon \over \sqrt{2\epsilon+\epsilon^2}} \int_0^{2\pi} \phi '' (\theta)  \log (1+ \epsilon - \sin\theta) d\theta
$$
Hence $\| \eta_\omega\|$, taken in $C^2(S^1)^\ast$, satisfies
$$
\| \eta_\omega\| \le {c \over  \sqrt \epsilon} \int_0^{2 \pi} | \log (1+\epsilon - \sin \theta )| d\theta
$$
for some constant $c$, and the integral above is bounded as $\epsilon \to 0$ (this follows from the fact that $\log x$ has an integrable singularity at $0$).  Hence $\eta$ is an element of $E$, and the proof is complete.
\qed

The eigenvector described above has an interesting interpretation.  As we saw in Section~4, if the density function $g$ is constant on $[0 , \omega_0]$,  the model with critical coupling $K_c = 2 / \pi g(0)$ has  a  family $\rho(t)$ of special positive states parametrized by $t \in (0, \omega_0]$.  The tangent vector $\eta$ described above is exactly the derivative of $\rho(t)$ with respect to $t$, at the value $t = KR$.
\b
\noindent {\bf Remark on $\bf \sigma(L)$ for the incoherent  state.}  All our methods apply as well to the incoherent state, and are in fact much easier to work through in this case. The details are in [Strogatz and Mirollo 1991], so we present here a brief summary of the results in the context of our current formulation.  There are no locked frequencies now, and the operator $M_\omega$ is just $-\omega D$ on $S^1$, which has spectrum $ i \omega {\Bbb Z}$.  So $M$ has purely continuous spectrum ${\Bbb R} i$, and the characteristic functions $h_c$ and $h_s$ are given by
$$
h_c(\lambda) = h_s(\lambda) = {1 \over 2}\int_{-\infty}^\infty {\lambda \over \lambda^2+\omega^2} \, g(\omega) d\omega.
$$
The only possible roots of the characteristic equation are real and positive, so the continuous spectrum of $L$ is also ${\Bbb R}i$.  If $\lambda > 0$, we express
$$
h_c(\lambda) =  \int_0^\infty {1 \over 1+ s^2} \, g(\lambda s)  ds
$$
to see that $h_c$ is strictly decreasing on $(0, \infty)$. Furthermore 
$$
\lim_{\lambda \to 0^+} h_c(\lambda) =  g(0) \int_0^\infty {1 \over 1+s^2} \, ds = {\pi g(0) \over 2} = K_c^{-1},
$$
so the characteristic equation $h_c(\lambda) = K^{-1}$  has a  root $\lambda > 0$ if and only if $K > K_c$, and hence the incoherent state is linearly unstable in this case.  It is of course no coincidence that the incoherent state loses stability at the same critical coupling at which the positive fixed states are born.

\b
\b
\b\noindent{\bf 9.  Concluding Remarks}
\b

We can now completely describe the spectrum of $L$ for all special positive states in the infinite-$N$ Kuramoto model (Figure 2).  For the fully locked case with $K > K_l$  the spectrum consists of the closed interval $[-KR, -\sqrt{(KR)^2-1}]$,  at most one negative eigenvalue in $[-\sqrt{(KR)^2-1}, 0)$, and  the eigenvalue at $0$ coming from  the rotational symmetry of the Kuramoto model.  This result is not surprising; the corresponding finite-$N$ model [Mirollo and Strogatz 2005] has stable fixed points for most choices of frequencies $\omega_i$ as $N \to \infty$, provided the coupling $K > K_l$.  The eigenvalues for these fixed points are all negative, and so $\sigma(L)$ is in some sense just the limit of the spectrum in the finite-$N$ case.

When we make the transition to the partially locked case at $K < K_l$, the spectrum explodes to include the entire imaginary axis, along with the segment $[-KR, 0]$, as shown in Figure 2c.  The presence of negative values in $\sigma(L)$ suggests some sort of asymptotic stability, but the presence of ${\Bbb R}i$ in the spectrum suggests more neutral behavior.  So in some sense what we have accomplished is to rule out any kind of exponential convergence to fixed states in the partially locked case.  

On the other hand, the order parameter $R(t)$ might still be able to approach its stationary value exponentially fast, due to a phase-mixing mechanism akin to Landau damping; this is known to occur for the incoherent state under some conditions [Strogatz et al.~1992; Balmforth and Sassi 2000]. One should also keep in mind that the partially locked states have only been shown to be {\it linearly} neutrally stable.  The small nonlinear terms neglected here could therefore prove crucial to understanding the full stability properties of the fixed states.  A careful nonlinear analysis along these lines, ideally one that is global in character, may be the next logical step in the ongoing attempt to make sense out of Kuramoto's marvelous calculation of 1975.  

\b
\b
\b\noindent{\bf Acknowledgments}
\b
Research supported in part by National Science Foundation grant DMS-0412757 to S.H.S.

\vfill\eject
\b
\b
\b\noindent{\bf References}
\b
J.~A. Acebron, L.~L. Bonilla, C.~J. Perez Vicente, F.~Ritort, R.~Spigler.  The Kuramoto model: a simple paradigm for synchronization phenomena. Reviews of Modern Physics 77, 137--185 (2005).
\s
D.~Aeyels and J.~A. Rogge.  Existence of partial entrainment and stability of phase locking behavior of coupled oscillators.  Progress of Theoretical Physics 112, 921--942 (2004).
\s
N.~J. Balmforth and R. Sassi.  A shocking display of synchrony.  Physica D 143, 21--55 (2000).
\s
L.~L. Bonilla, J.~C. Neu, and R. Spigler.  Nonlinear stability of incoherence and collective synchronization in a population of coupled oscillators. Journal of Statistical Physics 67, 313-330 (1992).
\s
N.~Chopra and M.~W. Spong.  On synchronization of Kuramoto oscillators. in: Proceedings of the 44th IEEE Conference on Decision and Control, and the European Control Conference 2005, Seville, Spain, pp.~3916--3922.
\s
J.~D. Crawford.  Amplitude expansions for instabilities in populations of globally coupled oscillators. Journal of Statistical Physics 74, 1047--1084 (1994).
\s
G.~B. Ermentrout.  Synchronization in a pool of mutually coupled oscillators with random frequencies.  Journal of Mathematical Biology 22, 1--9 (1985).
\s
A.~Jadbabaie, N.~Motee, M.~Barahona.  On the stability of the Kuramoto model of coupled nonlinear oscillators.  In: Proceedings of the American Control Conference, Boston, Massachusetts, 2004. 
\s
T. Kato.  Perturbation Theory for Linear Operators (Springer-Verlag, Berlin, 1995).
\s
I.~Z. Kiss, Y.~Zhai, J.~L.~Hudson. Emerging coherence in a population of chemical oscillators.  Science 296, 1676--1678 (2002). 
\s
Y.~Kuramoto.  Self-entrainment of a population of coupled nonlinear oscillators.  In: H.~Araki (Ed.), International Symposium on Mathematical Problems in Theoretical Physics, Lecture Notes in Physics, Vol. 39 (Springer, New York, 1975), pp. 420--422.
\s
Y.~Kuramoto. Chemical Oscillations, Waves, and Turbulence (Springer, Berlin,1984).
\s
S.~Lang.  Real and Functional Analysis  (Springer-Verlag, New York, 1993).
\s
Y.~Maistrenko, O.~Popovych, O.~Burylko, and P.~A. Tass.  Mechanism of desynchronization in the finite-dimensional Kuramoto model.  Physical Review Letters 93, 084102 (2004).
\s
R.~E. Mirollo and S.~H. Strogatz.  The spectrum of the locked state for the Kuramoto model of coupled oscillators.  Physica D 205, 249--266 (2005).
\s
Z.~Neda, E.~Ravasz, T.~Vicsek, Y.~Brechet, and A.~L. Barabasi.  Physics of the rhythmic applause.  Physical Review E 61, 6987--6992 (2000).
\s
J.~Pantaleone.  Synchronization of metronomes.  American Journal of Physics 70, 992--1000 (2002).
\s
J.~Pantaleone.  Stability of incoherence in an isotropic gas of oscillating neutrinos. Physical Review D 58, 073002 (1998).
\s
A.~Pikovsky, M.~Rosenblum, and J.~Kurths.  Synchronization: A Universal Concept in Nonlinear Sciences (Cambridge University Press, Cambridge, 2001).
\s
H.~Sakaguchi. Cooperative phenomena in coupled oscillator systems under external fields. Progress of Theoretical Physics 79, 39--46 (1988).
\s
H.~Sakaguchi and Y.~Kuramoto.  A soluble active rotator model showing phase transitions via mutual entrainment. Progress of Theoretical Physics 76, 576--581 (1986).
\s
S.~H. Strogatz.  From Kuramoto to Crawford: exploring the onset of synchronization in populations of coupled oscillators.  Physica D 143, 1--20 (2000).
\s
S.~H. Strogatz.  Sync: The Emerging Science of Spontaneous Order (Hyperion, New York, 2003).
\s
S.~H. Strogatz and R.~E. Mirollo.  Stability of incoherence in a population of coupled oscillators.  Journal of Statistical Physics 63, 613--635 (1991).
\s
S.~H. Strogatz, R.~E. Mirollo, and P.~C. Matthews.  Coupled nonlinear oscillators below the synchronization threshold: relaxation by generalized Landau damping.  Physical Review Letters 68, 2730--2733 (1992).
\s
S.~H. Strogatz, D.~M. Abrams, A. McRobie, B. Eckhardt, and E. Ott.  Crowd synchrony on the Millennium Bridge.  Nature 438, 43--44 (2005).
\s
J.~L. van Hemmen and W.~F. Wreszinski.  Lyapunov function for the Kuramoto model of nonlinearly coupled oscillators.  Journal of Statistical Physics 72, 145--166 (1993).
\s
C.~von Cube, S.~Slama, D.~Kruse, C.~Zimmermann, Ph.~W.~Courteille, G.~R.~M. Robb, N.~Piovella, and R.~Bonifacio.  Self-synchronization and dissipation-induced threshold in collective atomic recoil lasing. Physical Review Letters 93, 083601 (2004). 
\s
K. Wiesenfeld, P. Colet, and S.H. Strogatz.  Synchronization transitions in a disordered Josephson series array.  Physical Review Letters 76, 404--407 (1996).
\s
A.~T. Winfree.  Biological rhythms and the behavior of populations of coupled oscillators.  Journal of Theoretical Biology 16, 15--42 (1967).
\s
A.~T. Winfree.  The Geometry of Biological Time (Springer-Verlag, New York, 1980).

\vfill\eject

\noindent{\bf Figures}
\b
Figure~1: Order parameter $R$ for fixed states of the infinite-$N$ Kuramoto model, as a function of the coupling strength $K$.  Points $(K,R)$ on the curve $C$ correspond to special positive fixed states. Points $(K,R)$ below the curve and above the $K$-axis correspond to positive fixed states with nonzero weight functions $w$; all of these non-special states turn out to be unstable, as we remark at the end of Section 6. The curve $C$ has a vertical segment if and only if the density function $g$ is locally constant at $0$.  And when $\Omega$, the support of $g$,  is $[-1,1]$, points on or above the hyperbola $KR=1$ correspond to fully locked states, and points below correspond to partially locked states.  We present three qualitatively different cases: (a) $\Omega = [-1,1]$ and $ g(x) < g(0)$ for $x \ne 0$. (b) $\Omega = [-1,1]$ and $g$ is locally constant about its maximum.  Here $C$ has a vertical segment at $K=K_c$, corresponding to a 1-parameter family of special positive states, all with $K=K_c$ but different values of $R$. (c) $\Omega = \Bbb R$ and $ g(x) < g(0)$ for $x \ne 0$.  Full locking is never achieved.  
\b
Figure~2: The spectrum $\sigma(L)$ for the special positive fixed states.  (a) For fully locked states with $K>K_l$, $\lambda$ and $0$ are eigenvalues.  The rest of $\sigma(L)$ is continuous.  The zero eigenvalue follows from the rotational symmetry of the Kuramoto model. In contrast, the eigenvalue $\lambda$ is not present in all cases; it exists if and only if $h_c(-\sqrt{(KR)^2-1}) \ge K^{-1}$, as shown in the proof of Proposition 4. (b) For fully locked states at the bifurcation value $K=K_l$, the spectrum contains an eigenvalue at 0, and the rest of $\sigma(L)$ is continuous.  (c) For partially locked states with $K < K_l$, there is still an eigenvalue at 0.  The rest of the spectrum is continuous, as before, but now it includes the whole imaginary axis.  Hence the partially locked states are linearly neutrally stable.     
\vfill\eject

\epsfxsize4.5truein\epsfbox{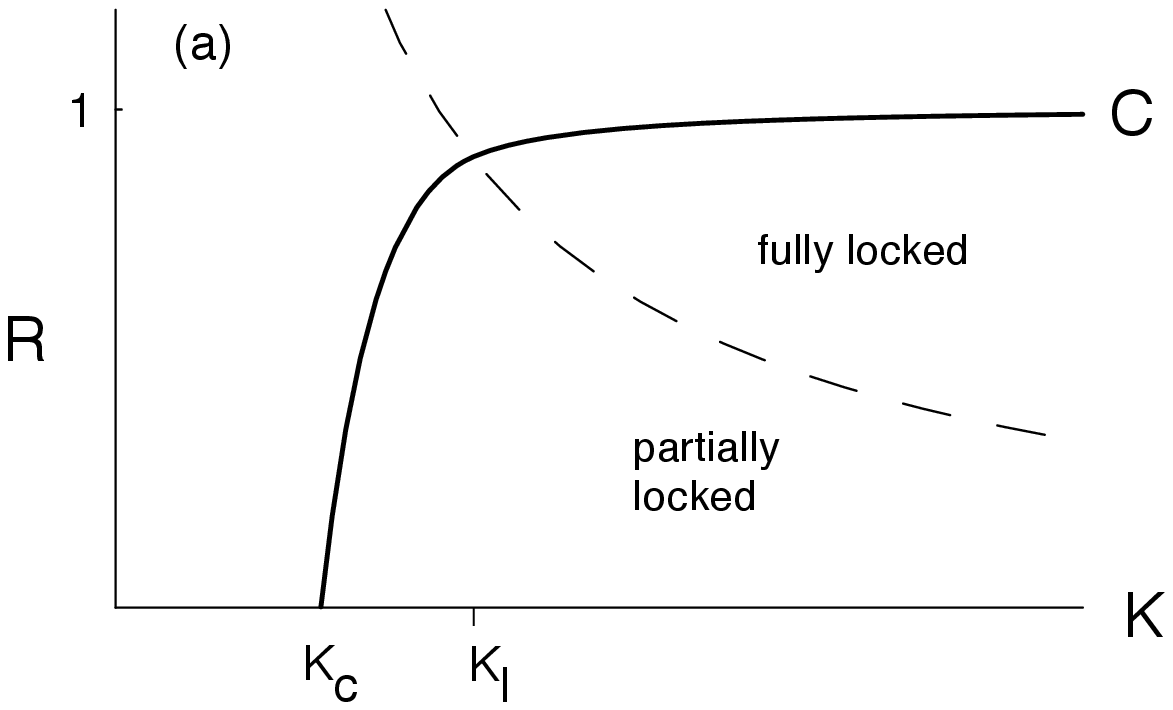}

\epsfxsize4.5truein\epsfbox{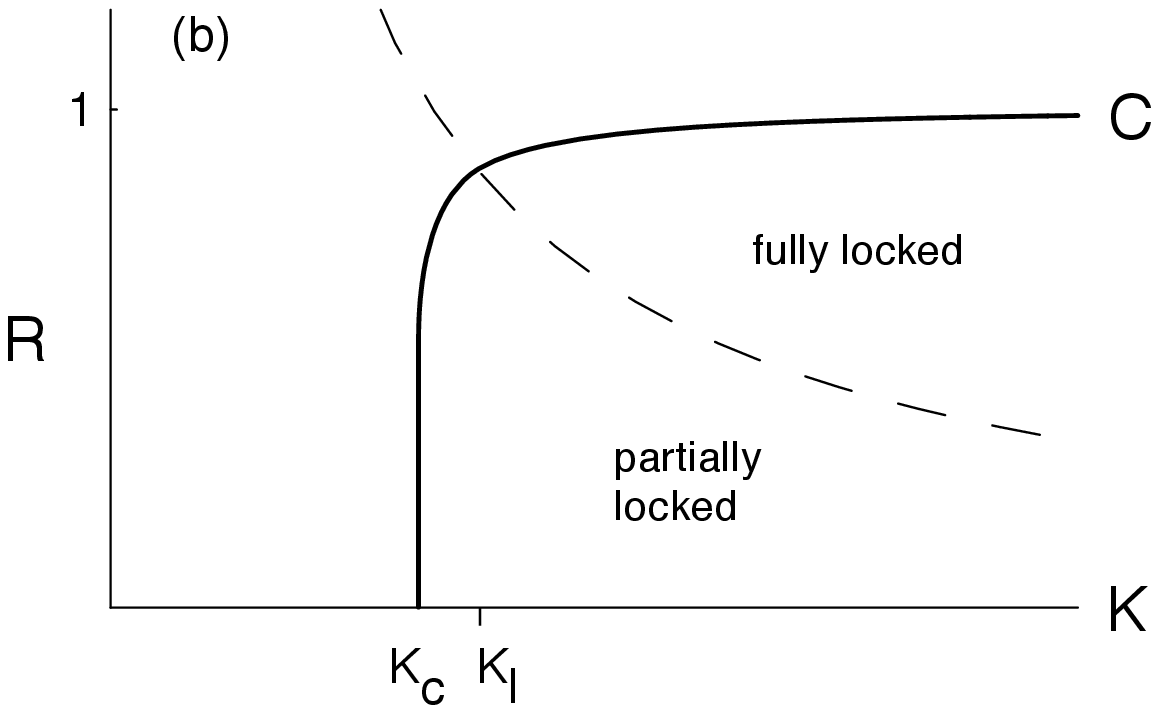}

\epsfxsize4.5truein\epsfbox{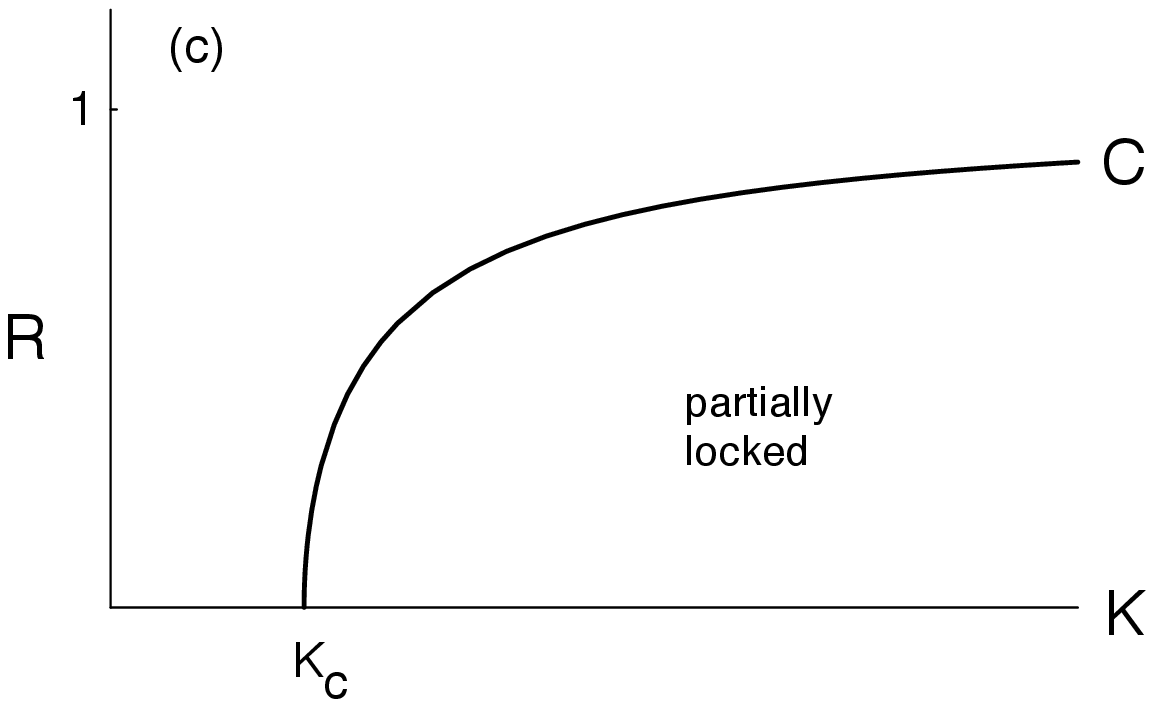}

\epsfxsize4.5truein\epsfbox{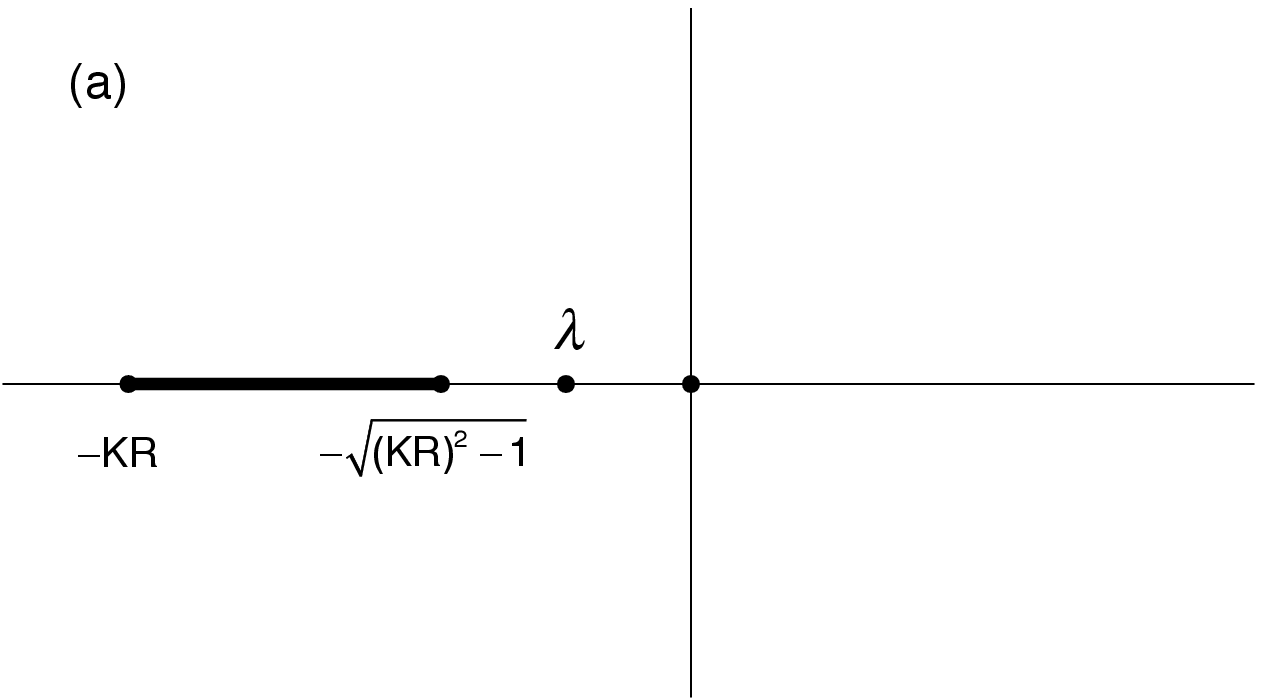}
\b

\epsfxsize4.5truein\epsfbox{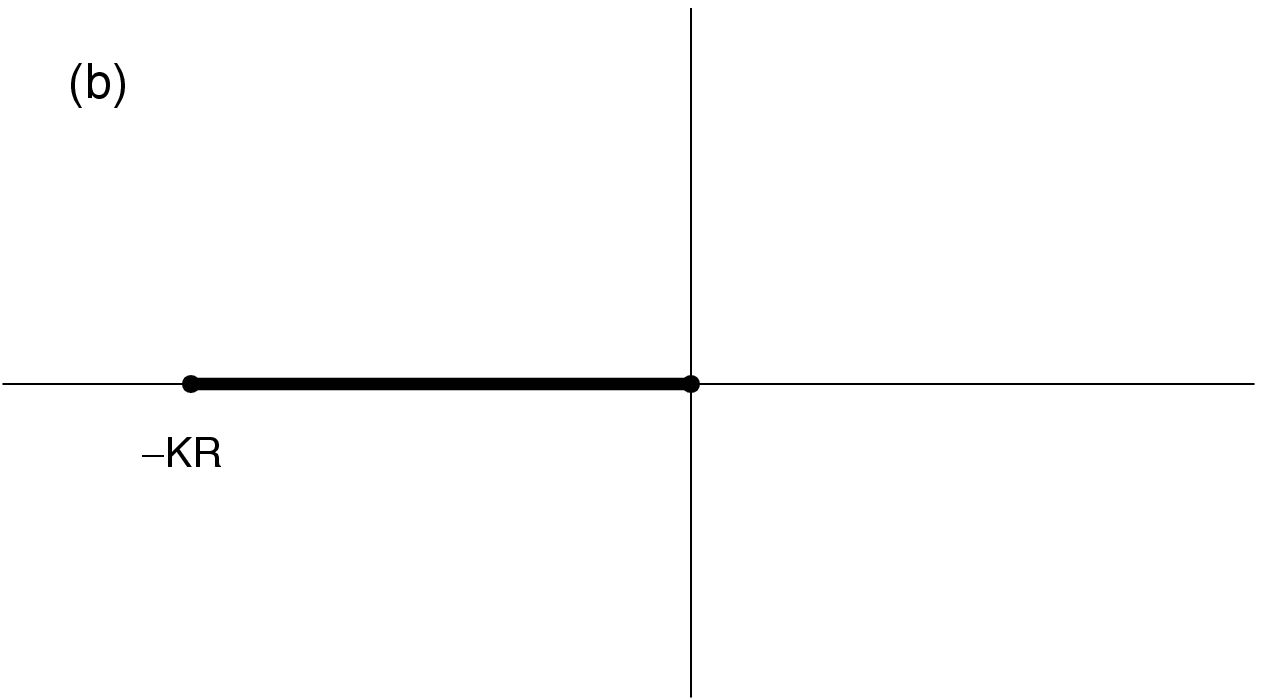}
\b

\epsfxsize4.5truein\epsfbox{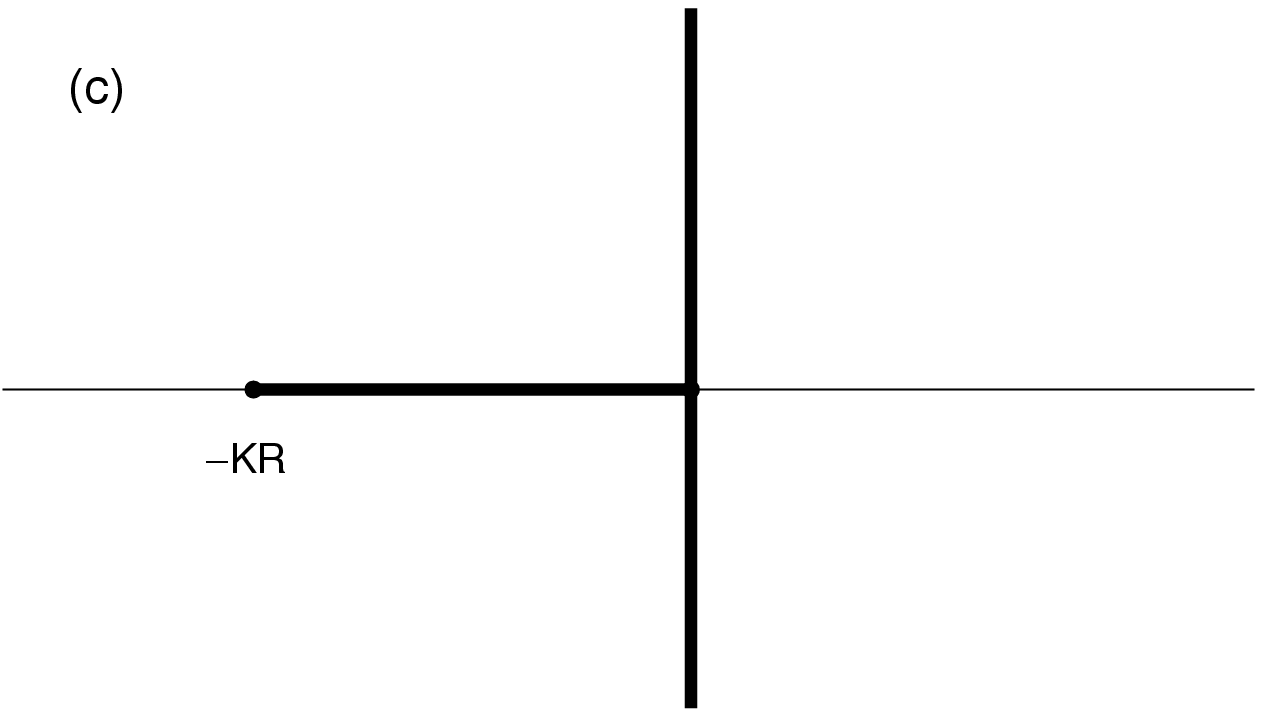}

 \bye